\documentclass[twocolumn, showpacs, preprintnumbers, superscriptaddress, aps, prl, floatfix, tightenlines]{revtex4-1}         
\pdfoutput=1
\usepackage{hyperref}
\usepackage{amsmath}
\usepackage{cleveref}
\usepackage{CJKutf8}
\usepackage{latexsym}
\usepackage{amstext}
\usepackage{amsfonts}
\usepackage{dsfont}
\usepackage{color}
\usepackage{amssymb}
\usepackage{relsize}
\usepackage{amsxtra}
\usepackage{verbatim}
\usepackage{hyperref}
\usepackage{graphics}
\usepackage{graphicx}
\usepackage{multirow}

\AtBeginDvi{\input{zhwinfonts}}
\renewcommand{\thetable}{\arabic{table}}

\begin{document}
\begin{CJK*}{UTF8}{shiai}

\title{High-temperature magnetism and crystallography of a YCrO$_3$ single crystal}

\author{Yinghao Zhu} 
\thanks{These two authors contributed equally.}
\affiliation{Joint Key Laboratory of the Ministry of Education, Institute of Applied Physics and Materials Engineering, University of Macau, Avenida da Universidade, Taipa, Macao SAR 999078, China.}
\author{Si Wu} 
\thanks{These two authors contributed equally.}
\affiliation{Joint Key Laboratory of the Ministry of Education, Institute of Applied Physics and Materials Engineering, University of Macau, Avenida da Universidade, Taipa, Macao SAR 999078, China.}
\author{Bao Tu} 
\affiliation{Joint Key Laboratory of the Ministry of Education, Institute of Applied Physics and Materials Engineering, University of Macau, Avenida da Universidade, Taipa, Macao SAR 999078, China.}
\affiliation{Department of Materials Science and Engineering, Shenzhen Key Laboratory of Full Spectral Solar Electricity Generation (FSSEG), Southern University of Science and Technology, No. 1088, Xueyuan Rd., Shenzhen 518055, Guangdong, China.}
\author{Shangjian Jin} 
\affiliation{State Key Laboratory of Optoelectronic Materials and Technologies, School of Physics and Engineering, Sun Yat-sen University, Guangzhou 510275, China.}
\author{Ashfia Huq} 
\affiliation{Oak Ridge National Laboratory, Oak Ridge, Tennessee 37831, USA.}
\author{J$\ddot{\textrm{o}}$rg Persson} 
\affiliation{J$\ddot{u}$lich Centre for Neutron Science JCNS-2 and Peter Gr$\ddot{u}$nberg Institut PGI, JARA-FIT, Forschungszentrum J$\ddot{u}$lich GmbH,
D-52425 J$\ddot{u}$lich, Germany.}
\author{Haoshi Gao} 
\affiliation{Joint Key Laboratory of the Ministry of Education, Institute of Applied Physics and Materials Engineering, University of Macau, Avenida da Universidade, Taipa, Macao SAR 999078, China.}
\affiliation{State Key Laboratory of Quality Research in Chinese Medicine, Institute of Chinese Medical Sciences (ICMS), University of Macau, Avenida da Universidade, Taipa, Macao SAR 999078, China.}
\author{Defang Ouyang} 
\affiliation{State Key Laboratory of Quality Research in Chinese Medicine, Institute of Chinese Medical Sciences (ICMS), University of Macau, Avenida da Universidade, Taipa, Macao SAR 999078, China.}
\author{Zhubing He} 
\affiliation{Department of Materials Science and Engineering, Shenzhen Key Laboratory of Full Spectral Solar Electricity Generation (FSSEG), Southern University of Science and Technology, No. 1088, Xueyuan Rd., Shenzhen 518055, Guangdong, China.}
\author{Dao-Xin Yao} 
\email{yaodaox@mail.sysu.edu.cn}
\affiliation{State Key Laboratory of Optoelectronic Materials and Technologies, School of Physics and Engineering, Sun Yat-sen University, Guangzhou 510275, China.}
\author{Zikang Tang} 
\email{zktang@um.edu.mo}
\affiliation{Joint Key Laboratory of the Ministry of Education, Institute of Applied Physics and Materials Engineering, University of Macau, Avenida da Universidade, Taipa, Macao SAR 999078, China.}
\author{Hai-Feng Li} 
\email{haifengli@um.edu.mo}
\affiliation{Joint Key Laboratory of the Ministry of Education, Institute of Applied Physics and Materials Engineering, University of Macau, Avenida da Universidade, Taipa, Macao SAR 999078, China.}

\date{\today}

\begin{abstract}

Magnetization measurements and time-of-flight neutron powder-diffraction studies on the high-temperature (300--980 K) magnetism and crystal structure (321--1200 K) of a pulverized YCrO$_3$ single crystal have been performed. Temperature-dependent inverse magnetic susceptibility coincides with a piecewise linear function with five regimes, with which we fit a Curie-Weiss law and calculate the frustration factor $f$. The fit results indicate a formation of magnetic polarons between 300 and 540 K and a very strong magnetic frustration. By including one factor $\eta$ that represents the degree of spin interactions into the Brillouin function, we can fit well the applied-magnetic-field dependence of magnetization. No structural phase transition was observed from 321 to 1200 K. The average thermal expansions of lattice configurations (\emph{a}, \emph{b}, \emph{c}, and \emph{V}) obey well the Gr$\ddot{\textrm{u}}$neisen approximations with an anomaly appearing around 900 K, implying an isosymmetric structural phase transition, and display an anisotropic character along the crystallographic \emph{a}, \emph{b}, and \emph{c} axes with the incompressibility $K^a_0 > K^c_0 > K^b_0$. It is interesting to find that at 321 K, the local distortion size $\Delta$(O2) $\approx$ 1.96$\Delta$(O1) $\approx$ 4.32$\Delta$(Y) $\approx$ 293.89$\Delta$(Cr). Based on the refined Y-O and Cr-O bond lengths, we deduce the local distortion environments and modes of Y, Cr, O1, and O2 ions. Especially, the Y and O2 ions display obvious atomic displacement and charge subduction, which may shed light on the dielectric property of the YCrO$_3$ compound. Additionally, by comparing Kramers Mn$^{3+}$ with non-Kramers Cr$^{3+}$ ions, it is noted that being a Kramers or non-Kramers ion can strongly affect the local distortion size, whereas, it may not be able to change the detailed distortion mode.

\end{abstract}

\maketitle
\end{CJK*}


\section{I. Introduction}

\begin{figure*} [!t]
\centering \includegraphics[width = 0.88\textwidth] {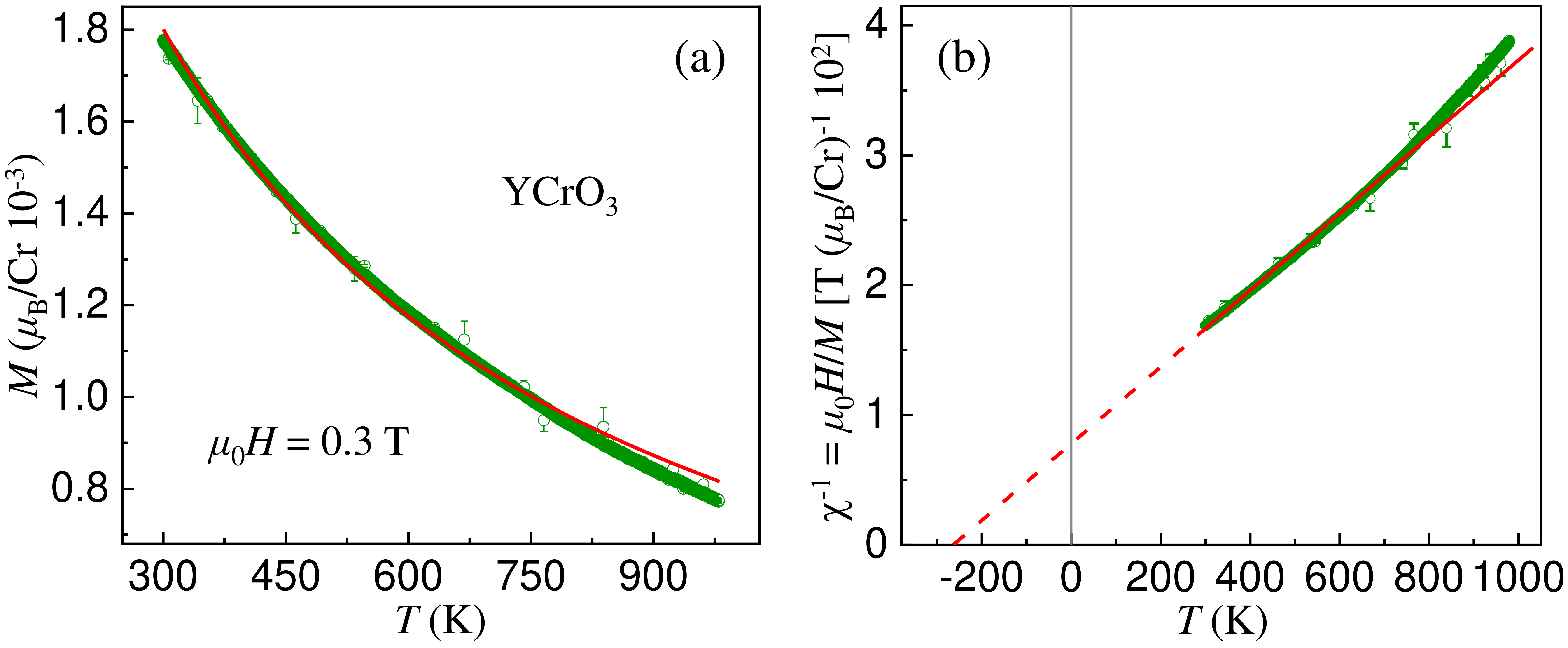}
\caption{(a) Measured magnetization (\emph{M}) of chromium ions in single-crystal YCrO$_3$ compound (circles) with an increase of temperature at $\mu_0H = $ 0.3 T ($\sim40008$ circles overlap each other so most of the error bars are embedded into the symbols). The solid line is a fit by Eq. (\ref{equation1}) as described in the text.
(b) Corresponding inverse magnetic susceptibility $\chi^{-1}$ (circles) of chromium ions in single-crystal YCrO$_3$ compound versus temperature. The solid line indicates a CW behavior of the data as described by Eq. (\ref{equation2}) from 300 to 980 K, which was extrapolated to $\chi^{-1} = 0$ (dashed line) to show the PM Curie temperatures $\theta_{\textrm{CW}}$.}
\label{Figure1}
\end{figure*}

The cooperating and competing interactions between charge, spin, lattice, and orbital degrees of freedom in condensed-matter science result in a variety of exotic macroscopic properties such as multiferroicity that continues to be an exciting field of research \cite{Fiebig2002, Kimura2003, Hur2004, Cheong2007, Kagawa2010, Valencia2011, Tokunaga2012}, urging experimental and theoretical scientists to unravel the complicated electron correlations and obtain a full understanding of the origin of the enduring phenomena.

Multiferroic materials are so called because multiple long-range orders such as magnetic and ferroelectric within them all coexist and{/}or interact with each other to give the substance extraordinary properties. They are potentially valuable materials for promising multifunctional magnetic and electric devices in information storage \cite{Cheong2007, Kagawa2010, Valencia2011, Tokunaga2012, Jones2014, Zhao2014} and provide a mode for information writing and reading \cite{Eerenstein2006, Nan2008, Dong2015, Spaldin2019}. Multiferroic materials can be classified into two types, based on the detailed couplings. In the first type, magnetism and ferroelectricity have different microscopic origins and thus a weak interaction existing between them. In such kinds of materials like YMnO$_3$, the ferroelectricity was usually caused by symmetry breaking \cite{Aken2004, Gibbs2011}. By contrast, in the second type of multiferroics, magnetism and ferroelectricity are strongly coupled with each other, and a particular magnetic order may lead to the appearance of ferroelectricity, like in the case of TbMnO$_3$ \cite{Kenzelmann2005}. As is well known, the magnetization in multiferroic materials is localized and originates from unpaired electrons in the partially filled \emph{d}- or \emph{f}-shell. Previous studies report many microscopic mechanisms for the formation of ferroelectricity such as charge order \cite{Choi2008}, lone electron pairs \cite{Hill200}, structural geometry \cite{Aken2004}, magnetoelectric coupling \cite{Mostovoy2006}, and oxygen vacancy \cite{Aschauer2013}.

The ABO$_3$ system with a perovskite-type crystalline structure constructs a significant part of the multiferroic materials. Among them, the BiFeO$_3$ thin film has been believed to be a typical representative in the research field of multiferroics. It simultaneously holds both a ferroelectric transition ($T_E$ = 1103 K) and a large remanent polarization ($P_\textrm{r}$ $\sim$ 55 $\mu$C/cm$^2$) above room temperature, which was attributed to the ordering of lone electron pairs \cite{Wang2003}. The YMnO$_3$ compound accommodates both a ferroelectric transition at $\sim$ 914 K and an antiferromagnetic (AFM) transition at $\sim$ 76 K \cite{Katsufuji2001}, and its ferroelectric polarization (\emph{P} $\sim$ 6 $\mu$C/cm$^2$) was suggested to be caused by the geometric variation, i.e., the tilting and distortion of MnO$_5$ blocks modifies the interaction between Y and O ions and thus forms the unbalanced dipoles of Y ions \cite{Aken2004, Fennie2005, Gibbs2011}. A very strong magnetoelectric effect exits in the TbMnO$_3$ compound and its charge polarization was induced by the specific magnetic order \cite{Kimura2003}. By changing chemical pressure, i.e., substituting the elements located at the crystallographic A or B site, a wide range of crystalline structures with potential multiferroic property could be realized \cite{Fedorova2018}.

Ferroelectricity is hard to coexist theoretically with magnetism in one perovskite so understanding the incompatible factors that are responsible for both properties has been a central topic in the area of multiferroics \cite{Cheong2007, Ederer2011, Balke2012}. A perovskite-type YCrO$_3$ compound was reported in 1954 \cite{Looby1954}. This compound is likewise a rare system to retain ferroelectricity and magnetism simultaneously. The YCrO$_3$ compound shows a magnetic phase transition at $T_\textrm{N} \approx$ 141.5 K; the applied-field dependent magnetization shows an unsaturated magnetic moment of mere $\sim$ 0.096 $\mu$$_\textrm{B}$ per Cr$^{3+}$ ion at 2 K and 7 T, which indicates that the YCrO$_3$ compound may behave like a canted AFM state below the \emph{T}$_\textrm{N}$; with regard to the low-temperature magnetic structure as well as the corresponding spin interaction parameters, the Dzyaloshinskii-Moriya interaction mechanism has been considered in a further report on our inelastic neutron scattering study of the YCrO$_3$ compound \cite{Jin2019}. A dielectric anomaly of the YCrO$_3$ compound was observed at $\sim$ 473 K with polarization \emph{P} $\sim$ 2 $\mu$C/cm$^2$, which was ascribed to the local Cr-ion off-centering displacement \cite{Serrao2005}. Electric polarization was reported in the orthorhombic GdCrO$_3$ polycrystalline samples below the canted AFM phase transition temperature \emph{T}$_\textrm{N}$ $\sim$ 167 K \cite{Rajeswaran2012}, which was accompanied by anomalies of both the Gd-Cr and Gd-O bond lengths \cite{Mahana2017, Mahana2018}. The ferroelectricity of the GdCrO$_3$ compound (\emph{P} $\sim$ 0.7 ${\mu}$C/cm$^2$) \cite{Rajeswaran2012} was proposed to be induced by the magnetic coupling between 4\emph{f} and 3\emph{d} electrons, as well as the atomic off-center displacement of a Gd site \cite{Mahana2017, Mahana2018}. It is pointed out that most of the previous research works on YCrO$_3$ compounds were focused on polycrystals due to the difficulty in growing large and high-quality single-crystalline samples, and the high-temperature magnetism and crystallographic information were seldom studied, though such kind of information is essential in completely understanding the previously observed dielectric anomaly \cite{Serrao2005}.

\begin{figure*} [!t]
\centering \includegraphics[width = 0.88\textwidth] {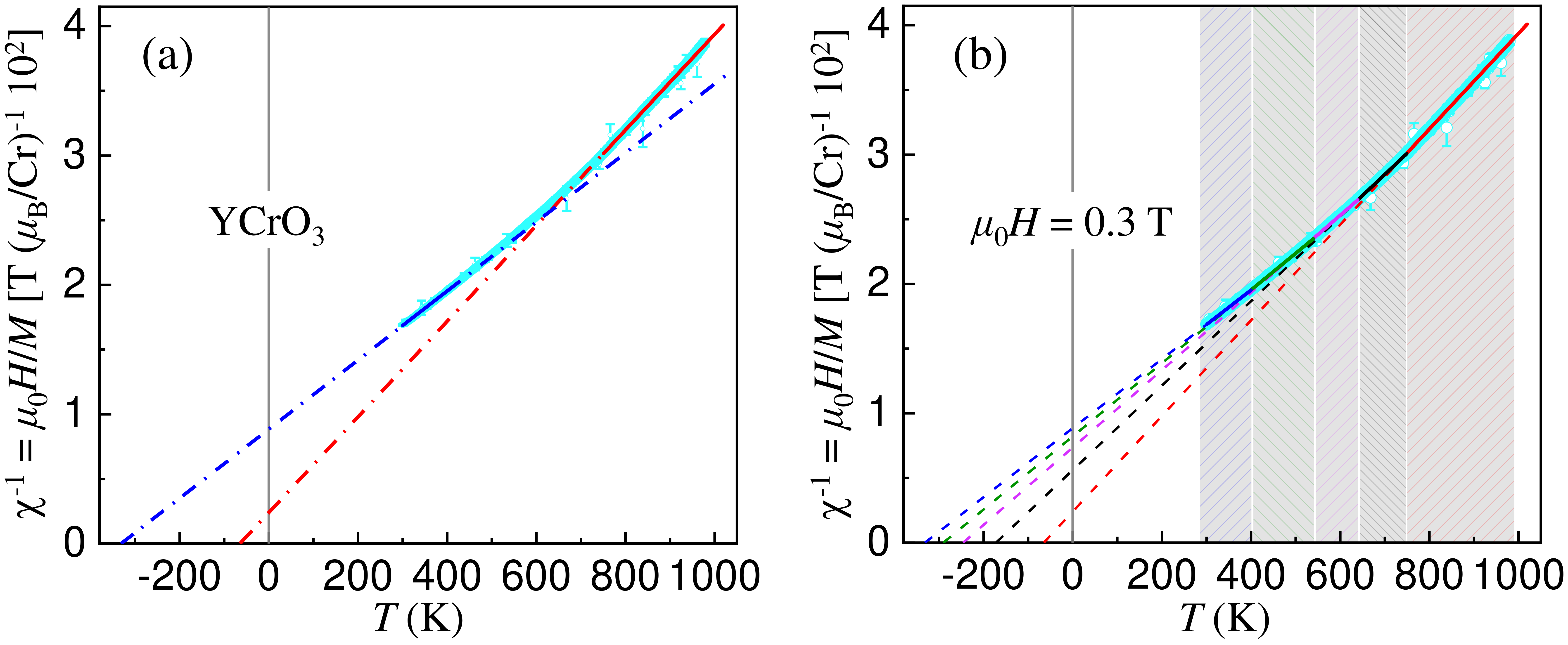}
\caption{(a) Inverse magnetic susceptibility $\chi^{-1}$ (circles) of chromium ions in single-crystal YCrO$_3$ compound versus temperature. The solid lines indicate CW behaviors of the data as described by Eq. (\ref{equation2}) at respective temperature regimes of 300--400 K and 750--980 K. They were extrapolated down to $\chi^{-1} = 0$ (dash-dotted lines) to show the PM Curie temperatures $\theta_{\textrm{CW}}$ and up to 1020 K (dash-dotted line). The fit results are listed in Table~\ref{Table1}.
(b) Inverse magnetic susceptibility $\chi^{-1}$ (circles) of chromium ions in single-crystal YCrO$_3$ compound versus temperature. The solid lines indicate CW behaviors of the data as described by Eq. (\ref{equation2}) at respective temperature regimes of 300--400 K, 400--540 K, 540--640 K, 640--750 K, and 750--980 K. They were extrapolated to $\chi^{-1} = 0$ (dashed lines) to show the PM Curie temperatures $\theta_{\textrm{CW}}$. The fit results are listed in Table~\ref{Table1}.}
\label{Figure2}
\end{figure*}

In this paper, the magnetization, crystal structure, thermal expansion, and local crystalline distortion of a pulverized YCrO$_3$ single crystal have been studied by PPMS DynaCool characterizations (300--980 K) and time-of-flight neutron powder-diffraction studies (321--1200 K). With our modified Brillouin function that includes a parameter $\eta$ representing the degree of magnetic interactions and a Curie-Weiss (CW) law, we uniquely determine the detailed magnetic parameters such as the effective paramagnetic (PM) moment, PM CW temperature, the parameter $\eta$, and the frustration factor \emph{f} to quantitatively understand the magnetism in YCrO$_3$ compounds. The space group of the crystal structure keeps $Pmnb$ in the entire temperature range. Anisotropic thermal expansion exists along the crystallographic \emph{a}, \emph{b}, and \emph{c} axes with the largest incompressibility $K_0$ along the $a$ axis, demonstrated by our fitting with the first-order Gr$\ddot{\textrm{u}}$neisen function taking into account only the phonon contribution for an insulator. The Y, O1, and O2 ions show very large local distortion size $\Delta$. We extract the detailed local distortion modes of Y, Cr, O1, and O2 ions. It is noted that distinct atomic displacement and a large charge subduction exiting for the Y and O2 ions are indicative of their important roles in producing the dielectric anomaly of YCrO$_3$ compound.

\section{II. Experimental}

Polycrystalline samples of YCrO$_3$ were prepared from stoichiometric mixtures of raw Y$_2$O$_3$ (ALFA AESAR, 99.9{\%}) and Cr$_2$O$_3$ (ALFA AESAR, 99.6{\%}) compounds by traditional solid-state reaction method \cite{Li2008}. After milling and mixing by a Vibratory Micro Mill (FRITSCH PULVERISETTE 0), the mixture was heated at 1000 $^{\circ}$C for 24 h with an increasing and decreasing temperature speed of 200 $^{\circ}$C/h in air to perform the process of pre-reaction. A similar heating procedure was carried out at 1100 $^{\circ}$C. After that, the resultant green mixture was isostatically pressed into a $\sim$ 12 cm cylindrical rod with a pressure of 70 MPa. The rod was then sintered once at 1300 $^{\circ}$C for 36 h in air. With above firing steps and milling and mixing with a ball of 50 mm in diameter after each heating process, we finally obtained a dense and homogenous pure polycrystalline YCrO$_3$ phase. High-quality single crystals of YCrO$_3$ were grown by the floating-zone (FZ) technique \cite{Li2008, Li2009} with a laser diode FZ furnace (Model: LD-FZ-5-200W-VPO-PC-UM) \cite{Zhu2019}.

The DC magnetization was measured with a temperature increasing speed of 1 K/min at 0.3 T in the temperature range from 300 to 980 K on a Quantum Design Physical Property Measurement System (PPMS DynaCool instrument). The \emph{M} curves at 300, 500, 700, and 900 K versus applied magnetic field ($\mu_0H$) up to 14 T were also recorded.

\begin{figure} [!t]
\centering \includegraphics[width = 0.48\textwidth] {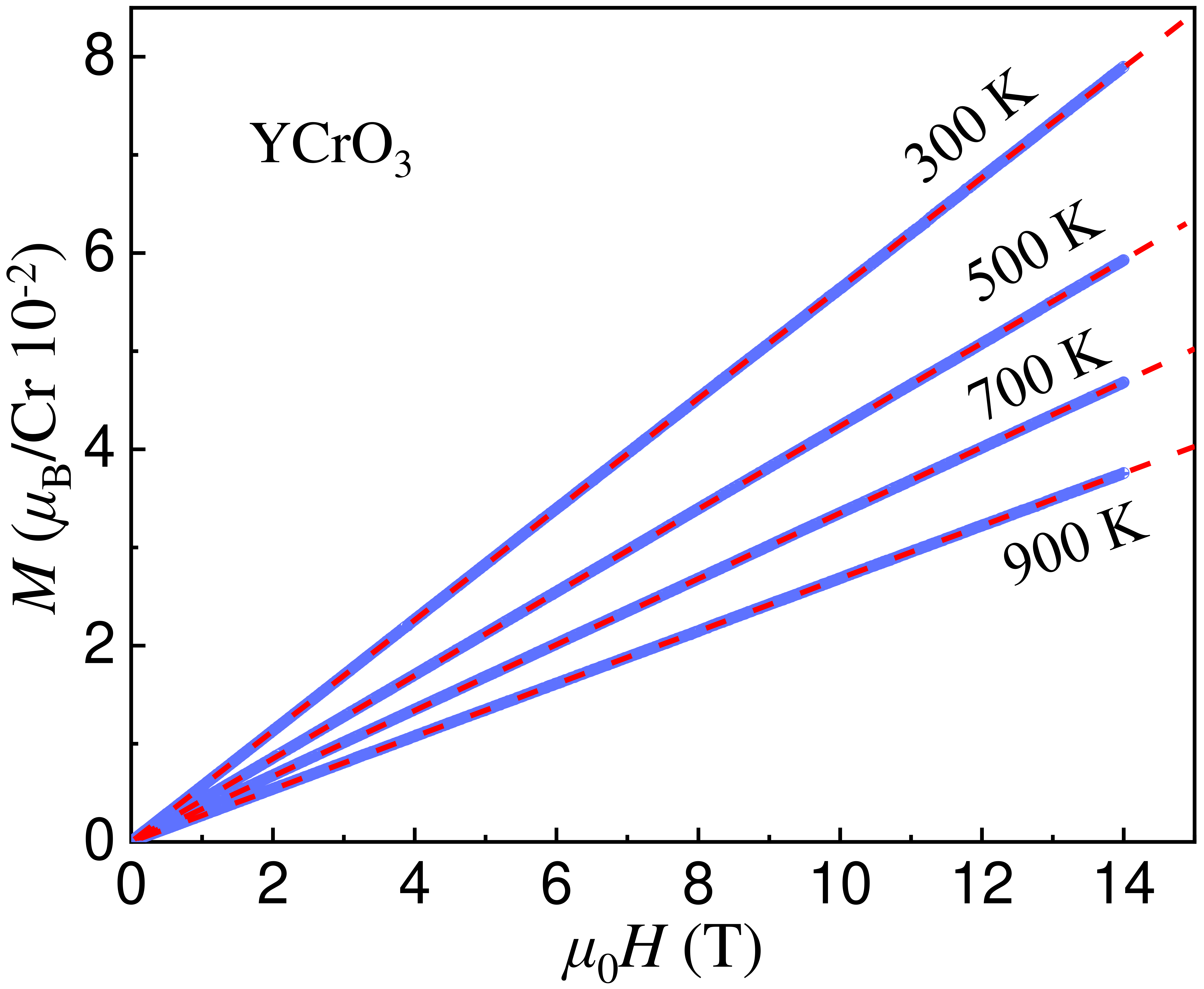}
\caption{Measured magnetization per chromium ion in single-crystal YCrO$_3$ compound (circles) as a function of applied magnetic fields up to 14 T at 300, 500, 700, and 900 K. The dashed lines are fits to Eq. (\ref{Br}). See detailed analysis in the text. Error bars are standard deviations and embedded into the circles because collected data points ($\sim$ 1600) at respective temperatures overlap each other.}
\label{Figure3}
\end{figure}

\begin{table}
\renewcommand*{\thetable}{\Roman{table}}
\caption{Theoretical quantum numbers for YCrO$_3$ compound: spin \emph{S} and the Land$\acute{\textrm{e}}$ factor $g_J$. We summarize the theoretical (theo) and measured (meas) (Fig.\ \ref{Figure1}) values of effective (eff) chromium moment, $\mu_{\textrm{eff}}$, and PM Curie temperature, $\theta_{\textrm{CW}}$. $R^2$ represents the goodness of fit. $T_\textrm{N} = 141.5$(1) K was extracted from our low-temperature (5--300 K) magnetization measurement. We also calculated the magnetic frustration factor \emph{f}. The $\eta$ factor is from Eq.~(\ref{Br}). See detailed analyses in the text. The numbers in parenthesis are the estimated standard deviations of the last significant digit.}
\label{Table1}
\begin{tabular} {llllll}
\hline
\hline
\multicolumn{6}{c} {A YCrO$_3$ single crystal}                                                                                                     \\
\hline
$S$                                                            &                           \multicolumn{5}{c} {3/2}                                \\
$g_J$                                                          &                           \multicolumn{5}{c} {2}                                  \\
$\mu_{\textrm{eff{\_}theo}}$ $(\mu_\textrm{B})$                &                           \multicolumn{5}{c} {3.873 }                             \\
\hline
\multicolumn{6}{c} {\emph{M} versus \emph{T} @ 0.3 T}                                                                                                     \\
\emph{T} (K)                                                   & 300--400      & 400--540       & 540--640      & 640--750       & 750--980         \\
$\mu_{\textrm{eff{\_}meas}}$ $(\mu_\textrm{B})$                & 4.09(1)       & 3.97(1)        & 3.86(1)       & 3.70(1)        & 3.47(1)         \\
$\theta_{\textrm{CW}}$ (K)                                     & --331.6(1)    & --290.3(1)     & --245.7(2)    & --172.0(2)     & --64.8(2)        \\
$f = |\theta_{\textrm{CW}}|/T_\textrm{N}$                      & 2.34(1)       & 2.05(1)        & 1.74(1)       & 1.22(1)        & 0.46(1)         \\
$R^2$                                                          & 0.999         & 0.999          & 0.999         & 0.999          & 0.999           \\
\hline
\multicolumn{6}{c} {\emph{M} versus $\mu_\textrm{0}H$ {@} \emph{T}}                                                                                                     \\
\emph{T} (K)                                                   & 300           & 500            & 700           & 900            &                 \\
$\eta$                                                         & 0.50(1)       & 0.63(1)        & 0.70(1)       & 0.72(1)        &                 \\
\hline
\hline
\end{tabular}
\end{table}

We got the powder sample for structural study by carefully pulverizing one as-grown YCrO$_3$ single crystal. High-resolution time-of-flight neutron powder-diffraction studies were carried out at the POWGEN diffractometer (SNS, USA) from 321 to 1200 K to explore the detailed temperature-dependent structural information. A large $Q$ ($= 2\pi/d$) band was measured from 0.759 to 3.695 {\AA}$^{-1}$. Neutrons are very sensitive to oxygen atoms in comparsion to x rays, and time-of-flight neutrons naturally overcome the $\lambda/2$ problem. Therefore, the band at high \emph{Q} regime is able to monitor all possible structural modifications. We use the FULLPROF SUITE \cite{Carvajal1993} to refine all collected time-of-flight neutron powder-diffraction data. The scale factor, lattice constants, zero epithermal shift, and background contribution determined using a linear interpolation between automatically selected 30 points, peak profile shape, atomic positions, isotropic thermal parameters, and preferred orientation were all refined in the final analysis step.

\begin{figure} [!t]
\centering
\includegraphics[width = 0.48\textwidth] {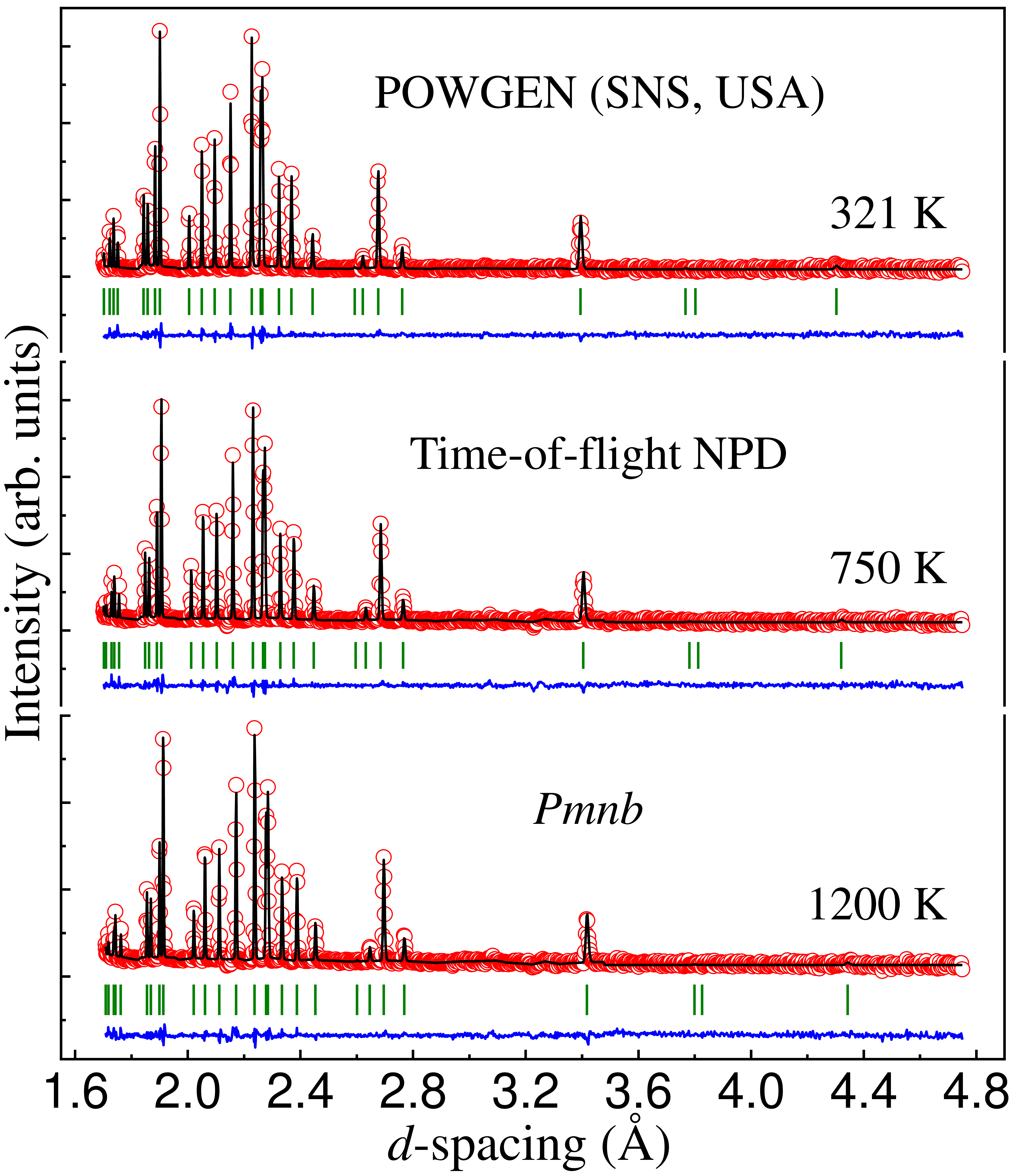}
\caption{Observed (circles) and calculated (solid lines) time-of-flight neutron powder-diffraction patterns of a pulverized YCrO$_3$ single crystal, collected on the POWGEN diffractometer (SNS, USA) at 321, 750, and 1200 K. The vertical bars mark the positions of nuclear Bragg reflections (space group $Pmnb$). The lower curves represent the difference between observed and calculated patterns.}
\label{Figure4}
\end{figure}

\begin{figure} [!t]
\centering
\includegraphics[width = 0.48\textwidth] {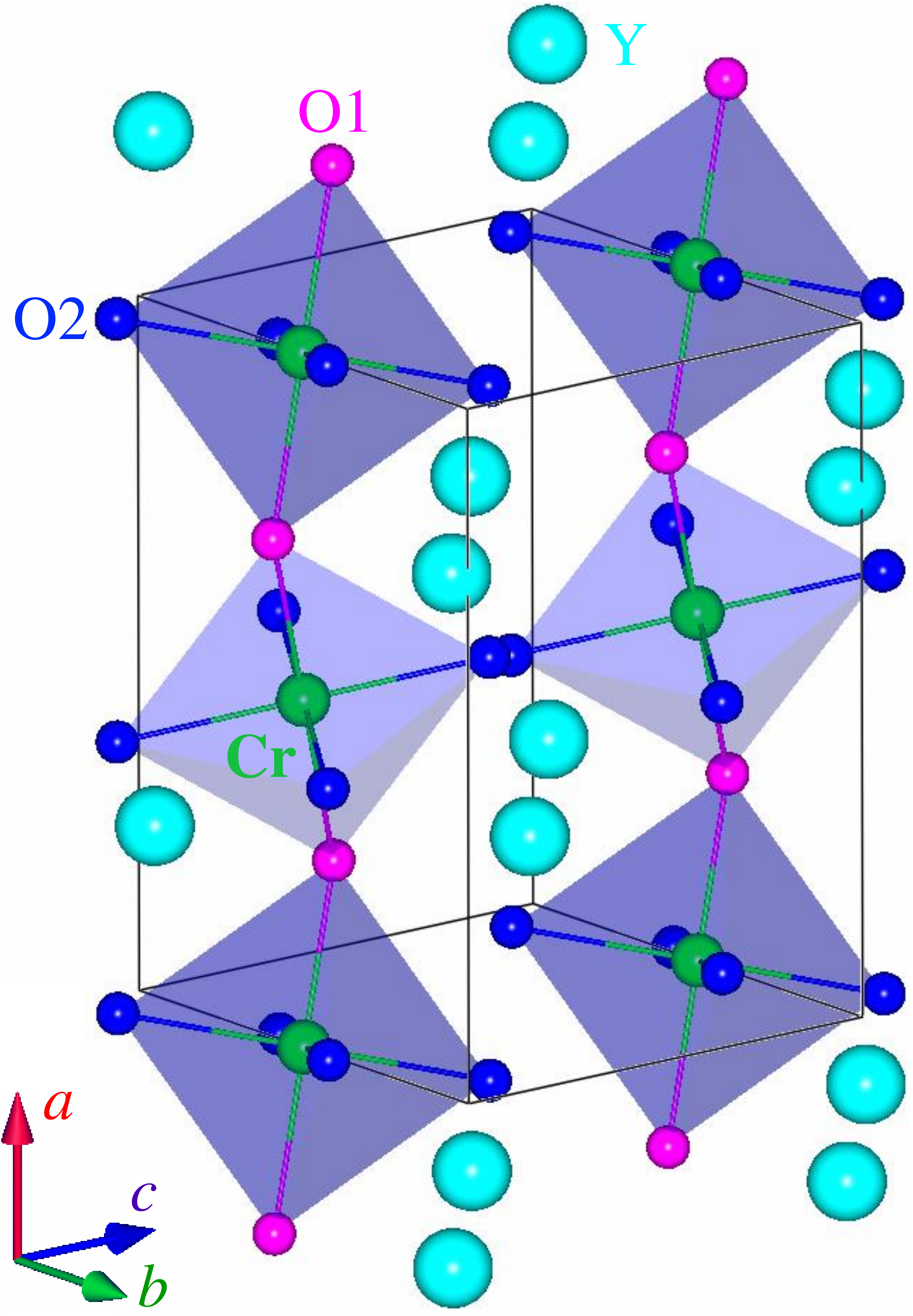}
\caption{Crystal structure (space group $Pmnb$) with one unit cell (solid line) of the YCrO$_3$ single crystal within the present experimental accuracy at the studied temperature regime of 321--1200 K. The solid balls were labeled as Y, Cr, O1, and O2 ions, respectively.}
\label{Figure5}
\end{figure}

\section{III. Results and discussion}

\subsection{A. Magnetization versus temperature}

\begin{table}
\renewcommand*{\thetable}{\Roman{table}}
\caption{Representative refined structural parameters, including lattice constants, unit-cell volume, atomic positions, isotropic thermal parameters (\emph{B}), bond lengths, bond angles, and distortion parameter ($\Delta$) of CrO$_6$ octahedra, of the pulverized single-crystal YCrO$_3$ compound by the FULLPROF SUITE \cite{Carvajal1993} with crystal structure (\emph{Pmnb}) as shown in Fig.\ \ref{Figure5} at 321, 750, and 1200 K. The Wyckoff site of each ion and the relevant reliability factors were also listed. The numbers in parenthesis are the estimated standard deviations of the last significant digit.}
\label{Table2}
\begin{ruledtabular}
\begin{tabular} {llll}
\multicolumn {4}{c}{Pulverized YCrO$_3$ single crystal}                                           \\
\multicolumn {4}{c}{(Orthorhombic, space group $Pmnb$, $Z = 4$)}                                  \\
\hline
$T$ (K)                          &321         &750          &1200                                 \\
\hline
$a$ ({\AA})                      &7.5332(3)   &7.5615(3)    &7.5976(3)                            \\
$b$ ({\AA})                      &5.5213(2)   &5.5280(2)    &5.5367(3)                            \\
$c$ ({\AA})                      &5.2418(2)   &5.2636(2)    &5.2921(2)                            \\
$\alpha (\beta, \gamma)$
$(^\circ)$                       &90          &90           &90                                   \\
$V$ ({\AA}$^3$)                  &218.02(1)   &220.02(2)    &222.61(2)                            \\
\hline
Y                                &4\emph{c}   &4\emph{c}    &4\emph{c}                            \\
\emph{x}                         &0.25        &0.25         &0.25                                 \\
\emph{y}                         &0.0689(8)   &0.0675(9)    &0.0649(10)                           \\
\emph{z}                         &--0.0169(6)  &--0.0155(7)   &--0.0142(8)                           \\
\emph{B} (\AA $^2$)              &4.4(13)     &5.7(17)      &5.4(19)                              \\
\hline
Cr                               &4\emph{b}   &4\emph{b}    &4\emph{b}                            \\
$(x, y, z)$                      &(0, 0, 0.5) &(0, 0, 0.5)  &(0, 0, 0.5)                          \\
\emph{B} (\AA $^2$)              &3.8(12)     &4.8(15)      &4.5(16)                              \\
\hline
O1                               &4\emph{c}   &4\emph{c}    &4\emph{c}                            \\
\emph{x}                         &0.25        &0.25         &0.25                                 \\
\emph{y}                         &0.4654(7)   &0.4660(9)    &0.4668(10)                           \\
\emph{z}                         &0.1053(8)   &0.1039(9)    &0.1017(11)                           \\
\emph{B} (\AA $^2$)              &3.9(14)     &5.2(18)      &5.1(19)                              \\
\hline
O2                               &8\emph{d}   &8\emph{d}    &8\emph{d}                            \\
\emph{x}                         &0.0545(4)   &0.0531(5)    &0.0523(6)                            \\
\emph{y}                         &0.3039(5)   &0.3039(5)    &0.3044(6)                            \\
\emph{z}                         &--0.3082(5)  &--0.3074(6)   &--0.3070(7)                           \\
\emph{B} (\AA $^2$)              &3.9(14)     &5.2(18)      &5.1(19)                              \\
\hline
Y-O11 (\AA)                      &2.232(5)    &2.238(6)     &2.250(7)                             \\
Y-O12 (\AA)                      &2.281(6)    &2.291(7)     &2.308(8)                             \\
Y-O21 (\AA) ($\times 2$)         &2.269(4)    &2.282(5)     &2.287(6)                             \\
Y-O22 (\AA) ($\times 2$)         &2.487(4)    &2.507(5)     &2.533(6)                             \\
$<$Y-O$>$ (\AA)                  &2.338(2)    &2.351(2)     &2.366(3)                             \\
\hline
Cr-O1 (\AA) ($\times 2$)         &1.972(1)    &1.977(1)     &1.983(2)                             \\
Cr-O21 (\AA) ($\times 2$)        &1.988(3)    &1.989(3)     &1.993(4)                             \\
Cr-O22 (\AA) ($\times 2$)        &1.999(3)    &2.003(3)     &2.010(3)                             \\
$<$Cr-O$>$ (\AA)                 &1.986(1)    &1.990(1)     &1.995(1)                             \\
\hline
$\angle$Cr-O1-Cr $(^\circ)$      &145.54(5)   &145.98(6)    &146.66(7)                            \\
$\angle$Cr-O2-Cr $(^\circ)$      &145.46(11)  &145.96(12)   &146.15(14)                           \\
\hline
$\Delta$(Cr) $(\times 10^{-4})$  &0.306       &0.285        & 0.329                               \\
\hline
$R_\textrm{p}$                   &7.83        &7.44         &6.88                                 \\
$R_\textrm{wp}$                  &8.12        &8.09         &7.58                                 \\
$R_\textrm{exp}$                 &4.49        &4.31         &4.06                                 \\
$\chi^2$                         &3.26        &3.52         &3.49                                 \\
\end{tabular}
\end{ruledtabular}
\end{table}

To explore possible magnetoelectric coupling effect \cite{Zheng2004, Lilienblum2015} in a single-crystal YCrO$_3$ compound, we performed a magnetization measurement at DC field of 0.3 T from 300 to 980 K. That broad temperature regime is far above the weak ferromagnetic (FM) transition temperature, $\sim$ 140 K, so we could ignore magnetic anisotropy, and covers the temperature point of $\sim$ 473 K, at which the dielectric anomaly was reported \cite{Serrao2005}. As shown in Fig.\ \ref{Figure1}(a), we transferred the measured magnetization into magnetic moment $\mu_\textrm{B}$ per Cr$^{3+}$ ion. With an increase of temperature, the magnetization decreases smoothly in the whole temperature range. The data collected in pure PM state can be fit to
\begin{eqnarray}
\emph{M} = \frac{m}{T - \theta_{\textrm{CW}}},
\label{equation1}
\end{eqnarray}
where \emph{m} is a constant, and $\theta_{\textrm{CW}}$ is the PM CW temperature. The diamagnetism of Cr$^{3+}$ ions is a temperature-independent constant, $\sim$ --1.1 $\times$ 10$^{-5}$ emu/mol \cite{Bain2008}, which is quite tiny so it can be neglected reasonably. Actually, the measured Cr$^{3+}$ magnetization at 980 K and 0.3 T is equal to 4.33(1) emu/mol in this study, by far larger than its diamagnetism. We tried to fit the measured data, shown as the solid line in Fig.\ \ref{Figure1}(a). This results in $\theta_{\textrm{CW}} = -264.0(1)$ K [Fig.\ \ref{Figure1}(b)], and the goodness of fit $\chi^2 = 68.629$. As shown in Fig.\ \ref{Figure1}, it is obvious that the fit is not comparable to the collected data, indicating that the CW temperature should have multiple components as a change in temperature. To confirm this, we included a constant, $M_{\textrm{Cr}}$, into the Eq. (\ref{equation1}) to express the temperature-independent ordered magnetism of Cr$^{3+}$ ions \cite{Yin2015, Dash2017}, though the YCrO$_3$ compound stays in a pure PM state from 300 to 980 K. With the modified equation, we fit our temperature-dependent magnetization data and the resultant $M_{\textrm{Cr}} =$ --2.329(2) emu/mol, much smaller than the Cr$^{3+}$ diamagnetism, which is physical nonsense.

The linear increase of the inverse magnetic susceptibility $\chi^{-1} = \frac{\mu_0H}{M}$ with temperature in the pure PM state obeys well the molar susceptibility by CW law \cite{Li2009, Li2014, Li2015},
\begin{eqnarray}
\chi^{-1}(T) = \frac{3k_B(T - \theta_{\textrm{CW}})}{N_A \mu^2_{\textrm{eff}}},
\label{equation2}
\end{eqnarray}
where $k_B$ = 1.38062 $\times$ 10$^{-23}$ J/K is the Boltzmann constant, $N_A$ = 6.022 $\times$ 10$^{23}$ mol$^{-1}$ is the Avogadro's number, and $\mu_{\textrm{eff}}$ = $g \mu_\textrm{B} \sqrt{J(J + 1})$ is the effective PM moment. Here $J = S$ for the YCrO$_3$ compound. As shown in Fig.\ \ref{Figure2}(a), we first fit the data in the temperature regimes of 300--400 K and 750--980 K (solid lines) and extrapolated them to the whole temperature range (dash-dotted lines). The two lines intersect at $T \sim$ 620 K and are not able to cover all features of the data. In addition, there exists no anomaly in crystallographic information at 620 K as discussed below. Therefore, it is evident that $\chi^{-1}$ versus \emph{T} obeys a piecewise linear function as shown in Fig.\ \ref{Figure2}(b). Only if we separated the inverse magnetic susceptibility into five regimes were we able to model all features and fit well the data by Eq. (\ref{equation2}). The corresponding fit results as well as the goodness of fit, $R^2$, are listed in Table~\ref{Table1}. With increasing the temperature ranges, the measured effective PM moment, $\mu_{\textrm{eff{\_}meas}}$, decreases from 4.09(1) (300--400 K) to 3.47(1) $\mu_\textrm{B}$ (750--980 K), whereas the corresponding PM CW temperature, $\theta_{\textrm{CW}}$, increases from --331.6(1) (300--400 K) to --64.8(2) K (750--980 K). In the temperature ranges of 300--400 K and 400--540 K, the measured effective PM moments $\mu_{\textrm{eff{\_}meas}} =$ 4.09(1) and 3.97(1) $\mu_\textrm{B}$, respectively. Both values are larger than the calculated theoretical one $\mu_{\textrm{eff{\_}theo}} =$ 3.873 $\mu_\textrm{B}$ supposing that all valence states are ionic and integer. The enhancement may be attributed to unquenched orbital angular momentum or the existence of local FM clusters with short-range spin interactions, i.e., magnetic polarons \cite{Li2009, Teresa1997, Nogues2001, Li2012}. In the temperature range 540--640 K, $\mu_{\textrm{eff{\_}meas}} =$ 3.86(1) $\mu_\textrm{B}$ is consistent with $\mu_{\textrm{eff{\_}theo}}$. At elevated temperature regimes of 640--750 and 750--980 K, $\mu_{\textrm{eff{\_}meas}}$ becomes smaller than $\mu_{\textrm{eff{\_}theo}}$. Based on these observations, it is deduced that the above hypothesis on the formation of magnetic polarons is more reasonable. The increase of temperature easily destroys magnetic interactions, i.e., the forming ground of magnetic polarons.

It is interesting that in the whole studied temperature range, all deduced PM CW temperatures, $\theta_{\textrm{CW}}$, are negative, indicating an AFM interaction, and largely deviated from the weak FM transition temperature $T_\textrm{N} = 141.5$(1) K that was obtained from our magnetization study on the same single crystal at low temperatures of 5--300 K. The coappearance of AFM and FM phenomena may indicate a canted AFM state with strong magnetic frustration that can be characterized by a frustration parameter $f = \frac{|\theta_{\textrm{CW}}|}{T_\textrm{N}}$. The larger (than 1) the value of $f$, the stronger the corresponding magnetic frustration is \cite{Ramirez2001}. As listed in Table\ \ref{Table1}, the \emph{f} values are all larger than 1 except for the one in the temperature range 750--980 K, implying an existence of strong magnetic frustration and a complex low-temperature magnetic structure.

The values of $f$ for the RMnO$_3$ compounds were reported to be 10.1 (YMnO$_3$), 10.3 (LuMnO$_3$), and 7.8 (ScMnO$_3$) \cite{Katsufuji2001}. Within these compounds, the Mn ions form a triangular lattice in the hexagonal structure, therefore, there exists a geometrical spin frustration. It is pointed out that the crystallographic structure of YCrO$_3$ compound doesn't accommodate any geometric frustration \cite{Ramirez1994, Diep2004, Shores2005, HFLi2014, Li2015}. The magnetic frustration in YCrO$_3$ compound originates from the observation of AFM and FM behaviors, which necessitates a determination of the detailed low-temperature magnetic structure.

\subsection{B. Magnetization versus applied magnetic field}

In this paper, we focus on the measurement of high-temperature ($\geq$ 300 K) magnetic properties. As shown in Fig.\ \ref{Figure3}, we monitored the magnetization as a function of applied magnetic field up to 14 T at 300, 500, 700, and 900 K with a small piece of randomly orientated single crystal ($\sim$ 14.91 mg). While increasing applied magnetic field, the magnetization increases linearly at all temperatures. The measured magnetic moment at 300 K and 14 T is 0.079(1) $\mu_\textrm{B}$/Cr, mere $\sim$ 2.63\% of the theoretical saturated (sat) Cr$^{3+}$ moment $\mu_{\textrm{sat{\_}theo}} = g_J S \mu_\textrm{B}=$ 3 $\mu_\textrm{B}$ (Table\ \ref{Table1}).

For theoretically isolated atoms, the change of magnetization with applied magnetic field at high temperatures obeys a Brillouin function given by \cite{Stephen2001}
\begin{eqnarray}
&&M(\mu_\textrm{0}H) = \eta M^{\textrm{sat}}_{\textrm{theo}} B_J(x), \ \ \ \ \ \ \ \textrm{with}            \nonumber \\
&&B_J(x) = \frac{2J+1}{2J} \textrm{coth}\mathlarger{\mathlarger{\mathlarger{(}}}\frac{2J+1}{2J}x\mathlarger{\mathlarger{\mathlarger{)}}}-\frac{1}{2J} \textrm{coth}\mathlarger{\mathlarger{\mathlarger{(}}}\frac{1}{2J}x\mathlarger{\mathlarger{\mathlarger{)}}},
\label{Br}
\end{eqnarray}
where $M^{\rm sat}_{\rm theo} = g_J J \mu_\textrm{B} =$ 3 $\mu_\textrm{B}$ is the theoretical value of the saturated mole moment, $J = \frac{3}{2}$ is the total angular momentum, and $x = \frac{g_JJ\mu_{\rm B}\mu_\textrm{0}H}{k_BT}$. Equations~(\ref{Br}) were used to fit the collected data as shown in Fig.\ \ref{Figure3} (dashed lines). The corresponding $\eta$ values at respective temperatures are listed in Table\ \ref{Table1}. When $\eta =$ 1, i.e., the theoretical case, there exists no magnetic exchange. With a decrease of temperature, spin interactions and the resultant magnetic order become possible. Since rotating and aligning spin moments of an antiferromagnet necessitate a very strong applied magnetic field, depending on detailed exchange parameters \cite{Tian2010, Toft2012, HFLi2016}, measured magnetization will deviate from the value of the saturation moment, leading to the $\eta$ value getting smaller than 1 and becoming smaller and smaller with an increase of AFM domains. As temperature decreases, the $\eta$ values decrease, in agreement with the increase of \emph{f} factors (Table\ \ref{Table1}).

\begin{figure} [!t]
\centering
\includegraphics[width = 0.48\textwidth] {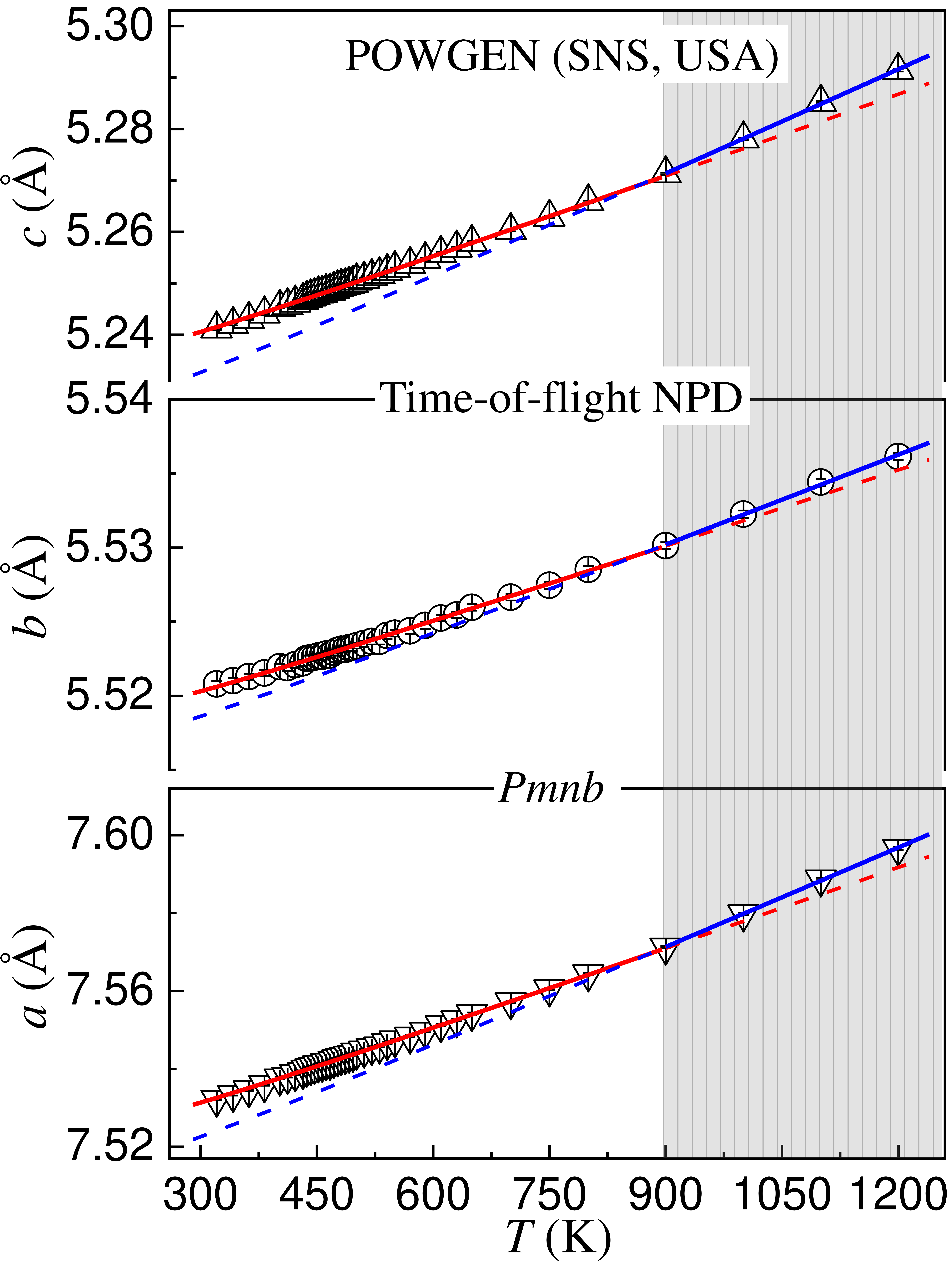}
\caption{Temperature dependence of the lattice constants, \emph{a}, \emph{b}, and \emph{c}, of the pulverized YCrO$_3$ single crystal (void symbols), which was extracted from our FULLPROF \cite{Carvajal1993} refinements based on the time-of-flight neutron powder-diffraction data collected on POWGEN diffractometer (SNS, USA) between 321 and 1200 K. The solid lines are theoretical estimates of the variation of structural parameters at the respective temperature regimes of 321--900 K and 900--1200 K, using the Gr$\ddot{\textrm{u}}$neisen model [Eqs. (\ref{Gr1}) and (\ref{Gr2})] with Debye temperature $\theta_\textrm{D}$ = 580 K, and extrapolated to the entire temperature range of 321--1200 K (dashed lines). Error bars are standard deviations obtained from our FULLPROF \cite{Carvajal1993} refinements in the \emph{Pmnb} symmetry.}
\label{Figure6}
\end{figure}

\begin{figure} [!t]
\centering
\includegraphics[width = 0.48\textwidth] {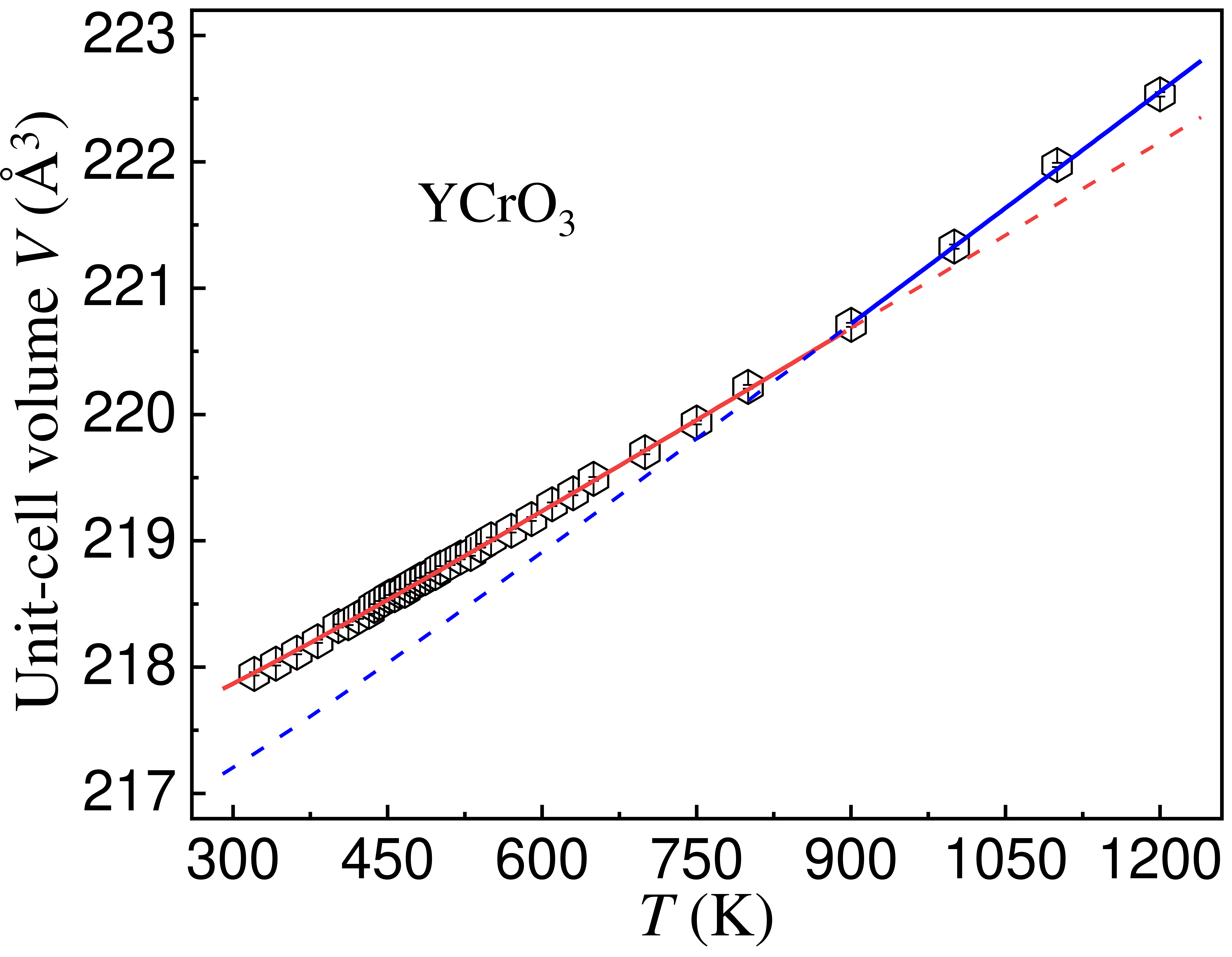}
\caption{Temperature-dependent unit-cell volume, \emph{V}, of the YCrO$_3$ single crystal (void symbols). This was extracted from our FULLPROF \cite{Carvajal1993} refinements based on the time-of-flight neutron powder-diffraction data collected on POWGEN diffractometer (SNS, USA) between 321 and 1200 K. The solid lines are theoretical estimates of the variation of \emph{V} within respective temperature regimes using the Gr$\ddot{\textrm{u}}$neisen model [Eqs. (\ref{Gr1}) and (\ref{Gr2})] with Debye temperature $\theta_\textrm{D}$ = 580 K and extrapolated to the whole temperature range (dashed lines). Error bars are standard deviations obtained from our FULLPROF \cite{Carvajal1993} refinements in the \emph{Pmnb} symmetry.}
\label{Figure7}
\end{figure}

\subsection{C. Time-of-flight neutron powder diffraction}

\begin{table}
\renewcommand*{\thetable}{\Roman{table}}
\caption{Fit parameters of the lattice configuration $\epsilon$ ($a, b, c$, and $V$) of YCrO$_3$ compound with Eqs.~(\ref{Gr1}) and (\ref{Gr2}), where $N =$ 5 and $k_B = 1.38062 \times 10^{-23}$ J/K, at the respective temperature regimes of 321--900 K and 900--1200 K [Figs.~\ref{Figure6} and \ref{Figure7}].
}
\label{Table3}
\begin{ruledtabular}
\begin{tabular} {lllll}
$T$ (K)                           & \multicolumn {2}{l}{321--900}                         & \multicolumn {2}{l}{900--1200}                                    \\
$\Theta_D$ (K)                    & \multicolumn {2}{l}{580}                             & \multicolumn {2}{l}{580}                                         \\
\hline
$\epsilon^a_0$ ({\AA})            & \multicolumn {2}{l}{7.52180(2)}                          & \multicolumn {2}{l}{7.51080(7)}                              \\
$\epsilon^b_0$ ({\AA})            & \multicolumn {2}{l}{5.51790(4)}                          & \multicolumn {2}{l}{5.51580(1)}                              \\
$\epsilon^c_0$ ({\AA})            & \multicolumn {2}{l}{5.23320(3)}                          & \multicolumn {2}{l}{5.22330(5)}                              \\
$\epsilon^V_0$ ({\AA}$^3$)        & \multicolumn {2}{l}{217.190(1)}                          & \multicolumn {2}{l}{216.350(9)}                              \\
\hline
$K^a_0$ ({\AA}/J)                 & \multicolumn {2}{l}{$3.380(7) \times 10^{17}$}           & \multicolumn {2}{l}{$4.160(5) \times 10^{17}$}               \\
$K^b_0$ ({\AA}/J)                 & \multicolumn {2}{l}{$8.390(1) \times 10^{16}$}           & \multicolumn {2}{l}{$9.910(8) \times 10^{16}$}               \\
$K^c_0$ ({\AA}/J)                 & \multicolumn {2}{l}{$2.590(4) \times 10^{17}$}           & \multicolumn {2}{l}{$3.300(8) \times 10^{17}$}               \\
$K^V_0$ ({\AA}$^3$/J)             & \multicolumn {2}{l}{$2.400(7) \times 10^{19}$}           & \multicolumn {2}{l}{$3.000(2) \times 10^{19}$}               \\
\end{tabular}
\end{ruledtabular}
\end{table}

To explore possible structural phase transitions of YCrO$_3$ compounds above room temperature, we carried out a time-of-flight neutron powder-diffraction study from 321 to 1200 K. Three representative neutron powder-diffraction patterns are shown in Fig.\ \ref{Figure4}. We carefully checked the Bragg peak shape, especially for the peaks located in the low-\emph{d} regime. All collected Bragg peaks were well indexed with the space group \emph{Pmnb} (orthorhombic structure). No additional peak and possible peak splitting were observed. We therefore conclude that within the present experimental accuracy, there exists no structural phase transition for the YCrO$_3$ compound from 321 to 1200 K. The refined crystal structure in one unit cell was depicted in Fig.\ \ref{Figure5}. The refined structural parameters at 321, 750, and 1200 K, including lattice constants, unit-cell volume, atomic positions, and thermal parameters are all listed in Table\ \ref{Table2}, where the extracted bond lengths, bond angles, and the distortion parameter of CrO$_6$ octahedra are also displayed. The low values of the reliability factors validate the goodness of our refinements.

\subsection{D. Anisotropic thermal expansion}

\begin{figure} [!t]
\centering
\includegraphics[width = 0.48\textwidth] {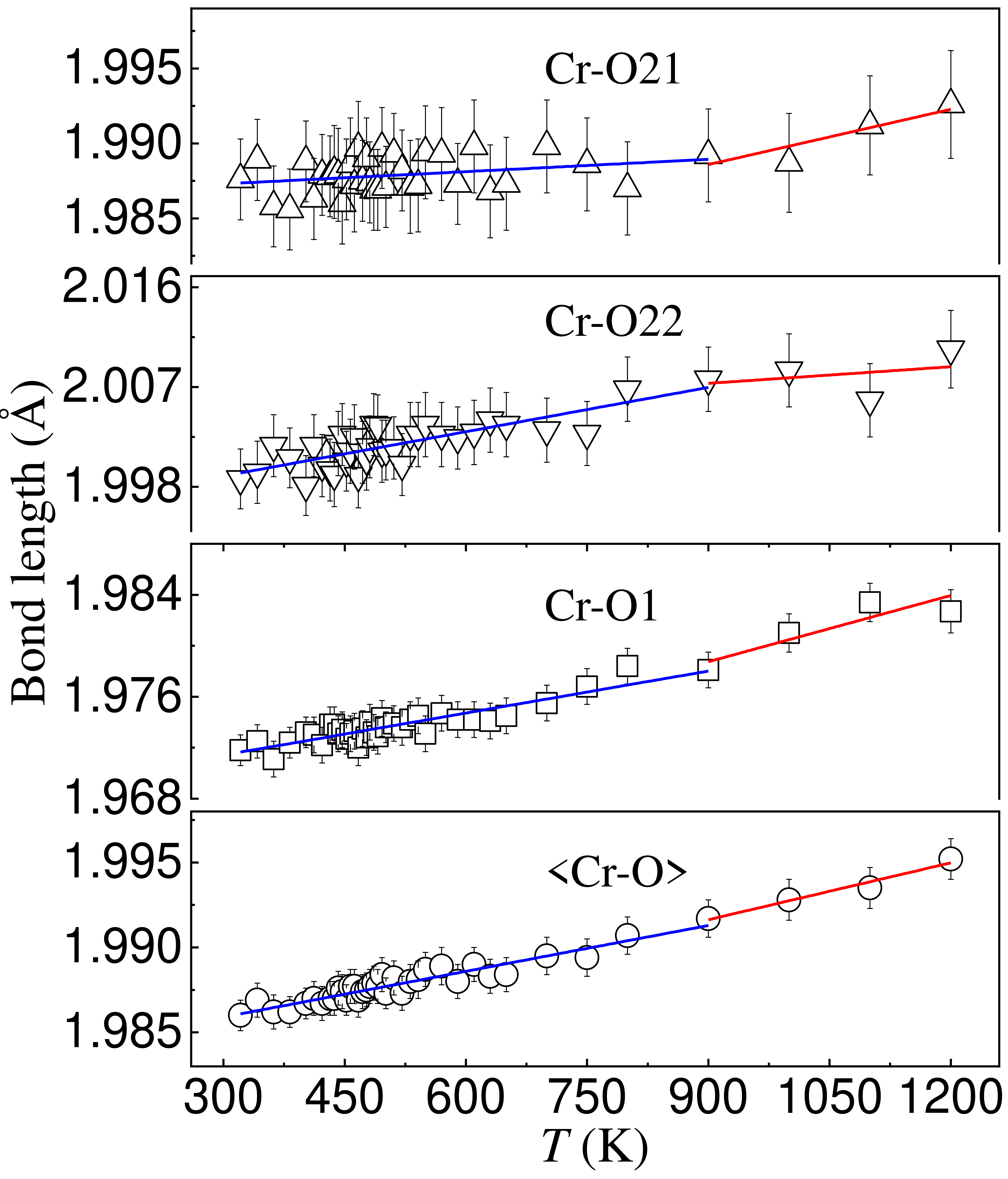}
\caption{Temperature-dependent bond lengths of Cr-O1, Cr-O21, and Cr-O22 as well as the averaged bond length of Cr-O, i.e., $<$Cr-O$>$, of the single-crystal YCrO$_3$ compound (void symbols). This was extracted from our time-of-flight neutron powder-diffraction study. Error bars are standard (for the Cr-O1, Cr-O21, and Cr-O22 bond lengths)/combined (for the Cr-O bond length) deviations. Solid lines are linear fits.}
\label{Figure8}
\end{figure}

\begin{figure} [!t]
\centering
\includegraphics[width = 0.48\textwidth] {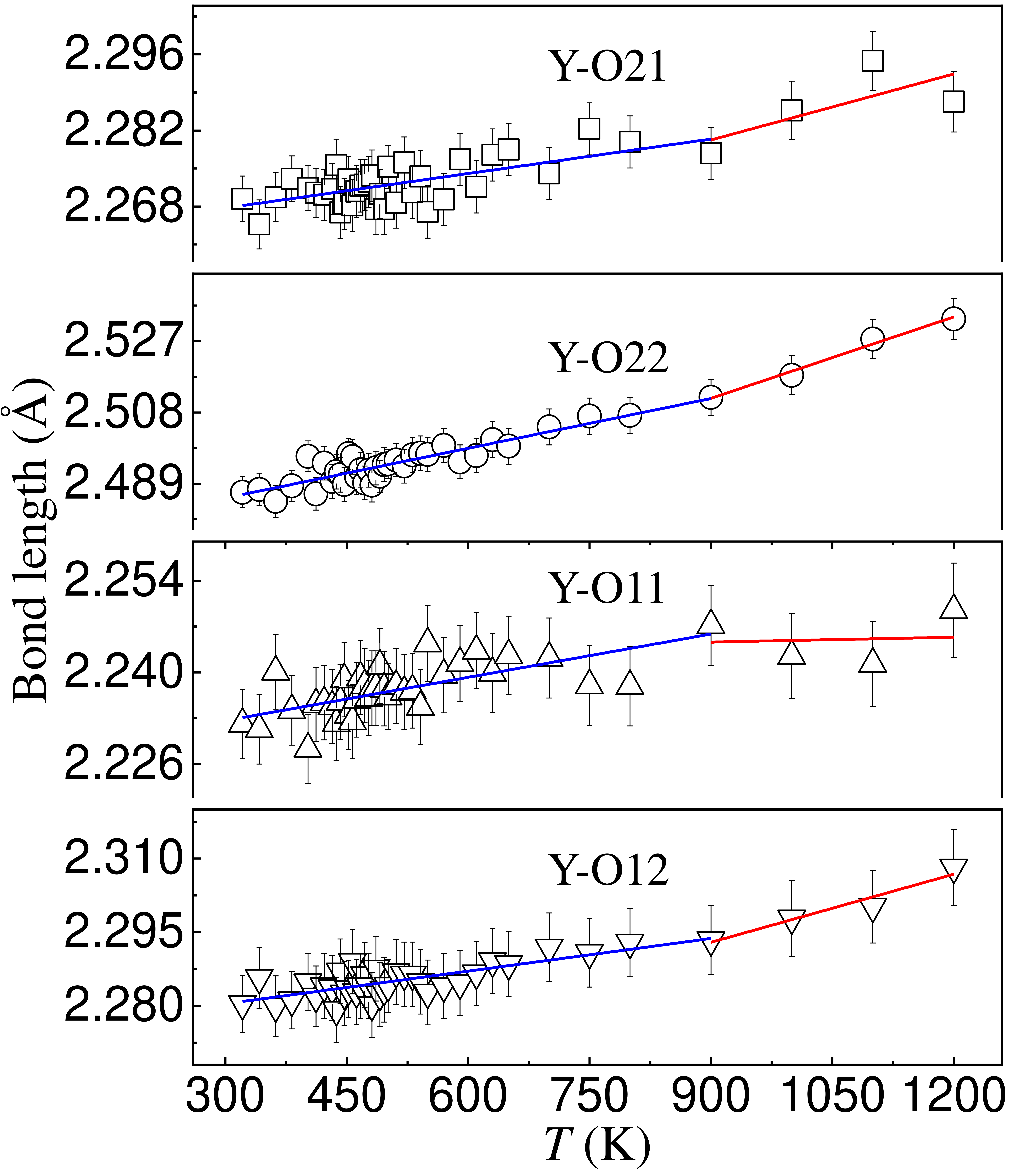}
\caption{Lengths of Y-O11, Y-O12, Y-O21, and Y-O22 bonds of the single-crystal YCrO$_3$ compound versus temperature varying from 321 to 1200 K (void symbols), which was extracted from our time-of-flight neutron powder-diffraction study. Error bars are standard deviations. Solid lines are linear fits.}
\label{Figure9}
\end{figure}

\begin{figure} [!t]
\centering
\includegraphics[width = 0.48\textwidth] {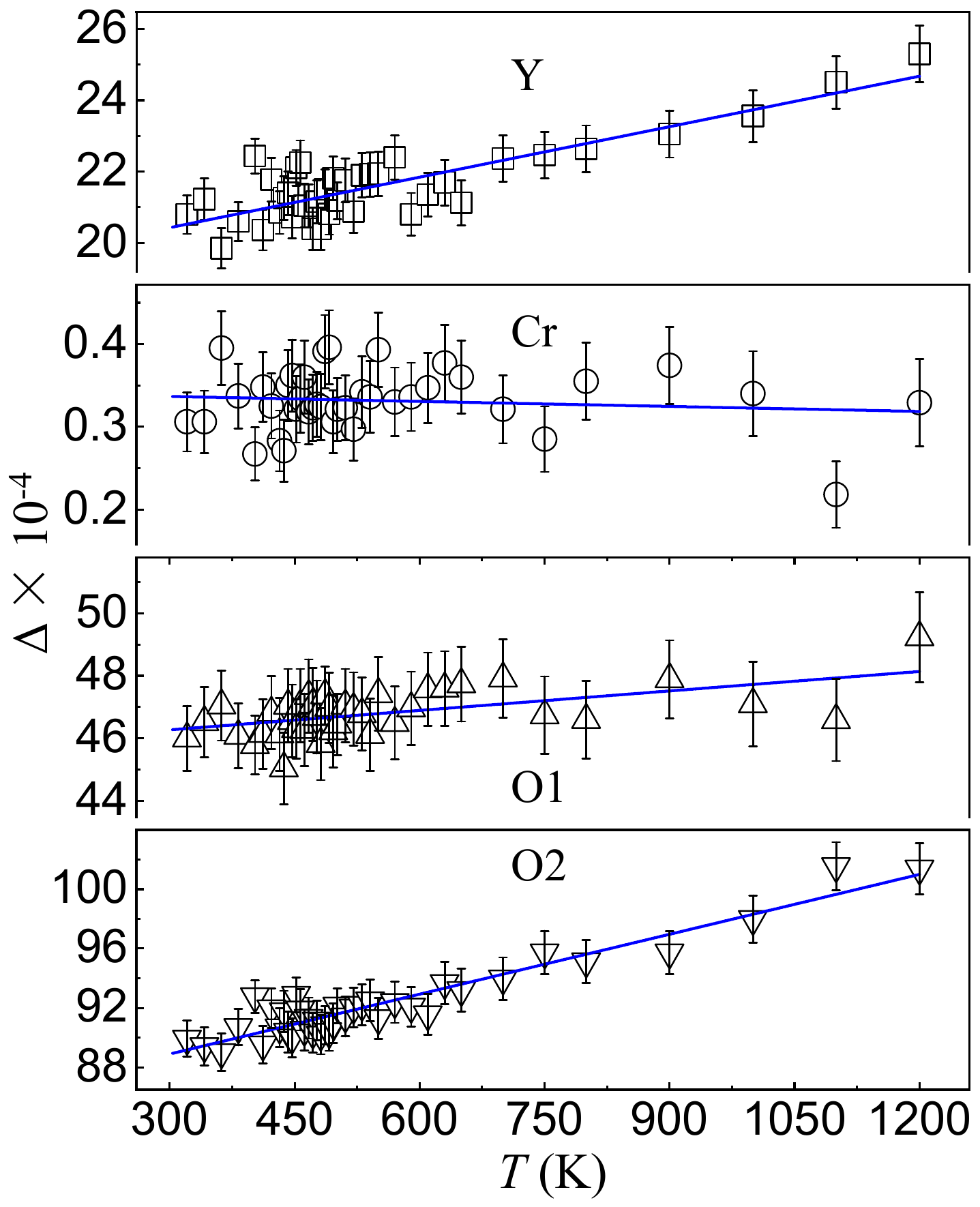}
\caption{Temperature variation of the distortion parameter $\Delta$ of Y, Cr, O1, and O2 ions of the single-crystal YCrO$_3$ compound (void symbols), calculated by Eq.~(\ref{distortion}) from the refined structural parameters between 321 and 1200 K. The error bar was estimated based on the propagation law of errors \cite{Zhu2019-E}. The solid lines are tentative linear fits.}
\label{Figure10}
\end{figure}

Figure\ \ref{Figure6} shows our refined lattice constants, \emph{a}, \emph{b}, and \emph{c}, at all temperatures from 321 to 1200 K. The corresponding change in unit-cell volume (\emph{V}) was depicted in Fig.\ \ref{Figure7}. For the insulating YCrO$_3$ compound, the contribution of lattice vibrations to the thermal expansion of the lattice configuration ($\varepsilon$) is much larger than that of electrons. Therefore, the temperature-dependent nonmagnetic contribution component of the thermal expansion is mainly from phonons. This can approximately be calculated based on the Gr$\ddot{\textrm{u}}$neisen function at zero pressure with the first-order fashion \cite{Li2012, Vocadlo2002, Wallace1998},
\begin{eqnarray}
\varepsilon(T) = \varepsilon_0 + K_0U,
\label{Gr1}
\end{eqnarray}
where $\varepsilon_0$ is the lattice configuration at 0 K, $K_0$ is a constant that reflects the incompressibility of the compound, and the internal energy \emph{U} can be estimated based on the Debye approximations,
\setlength\arraycolsep{1.4pt} 
\begin{eqnarray}
U(T) = 9Nk_BT\mathlarger{\mathlarger{\mathlarger{(}}}\frac{T}{\Theta_D}\mathlarger{\mathlarger{\mathlarger{)}}}^3 \int^{\frac{\Theta_D}{T}}_0 \frac{x^3}{e^x - 1}dx,
\label{Gr2}
\end{eqnarray}
where \emph{N} = 5 is the number of atoms per formula unit, and $\Theta_D =$ 580 K is the Debye temperature that can be determined by the upturn point of the $\varepsilon - T$ curve \cite{Li2012}. With the above Eqs.~(\ref{Gr1}) and (\ref{Gr2}), we fit the lattice configurations (\emph{a}, \emph{b}, \emph{c}, and \emph{V}) of YCrO$_3$ compound in two temperature ranges of 321--900 K and 900--1200 K (solid lines) and extrapolated the fits to the whole temperature regime (dashed lines) as shown in Figs.~\ref{Figure6} and \ref{Figure7}. The fit results are shown in Table\ \ref{Table3}. Within each temperature regime, as a whole, the temperature-dependent lattice configurations agree well with the theoretical estimations. However, at the boundary of the two temperature regimes, i.e., around 900 K, there exists an anomaly.

For lattice constants \emph{a}, \emph{b}, and \emph{c}, the incompressibility $K_0$ in the temperature range 900--1200 K is larger than that in $T =$ 321--900 K, which may be attributed to more developed phonon modes above 900 K. In both temperature regimes, $K^a_0 > K^c_0 > K^b_0$ indicates an anisotropic thermal expansion along the three crystallographic directions. This jointly results in the unit-cell volume expansion as shown in Fig.~\ref{Figure7}. ($K^a_{0(900{-}1200\textrm{K})}$ -- $K^a_{0(321{-}900\textrm{K})}$)/($K^a_{0(321{-}900\textrm{K})}$) = 22.71{\%}, ($K^b_{0(900{-}1200\textrm{K})}$ -- $K^b_{0(321{-}900\textrm{K})}$)/($K^b_{0(321{-}900\textrm{K})}$) = 18.24{\%}, and ($K^c_{0(900{-}1200\textrm{K})}$ -- $K^c_{0(321{-}900\textrm{K})}$)/($K^c_{0(321{-}900\textrm{K})}$) = 27.80{\%}, which implies that the temperature-dependent relative increase of $K_0$ is the largest one along the crystallographic \emph{c} axis.

For the YMnO$_3$ compound, a clear structural phase transition with a change in space group from centrosymmetric $P6_3/mmc$ to $P6_3cm$ occurs at 1258(14) K \cite{Jeong2007, Gibbs2011, Howard2013}. This structural phase transition temperature is much higher than its improper ferroelectricity transition temperature around 914 K and its AFM transition temperature around 76 K \cite{Katsufuji2001}. Clear anomalies were observed in the lattice constants \emph{a} and \emph{c} and unit-cell volume \emph{V} as well as the distance between Y1 and Y2 ions, the displacements of O3 and O4 ions, the tilting angle of apical O1-Mn-O2, and the lengths of Y1-O bonds, accompanying the structural phase transition \cite{Gibbs2011}. Moreover, an isosymmetric phase transition was found at $\sim$ 900 K, accompanied by a sharp decrease in polarization and anomalies in physical properties, which was attributed to a Y-O hybridization \cite{Gibbs2011}. In our study, for the YCrO$_3$ compound, structural anomalies were observed in lattice configurations (\emph{a}, \emph{b}, \emph{c}, and \emph{V}) as well as Cr-O and Y-O bond lengths around 900 K (as shown below). However, no clear change in the space group was distinguished. The preservation of the $Pmnb$ space group of YCrO$_3$ compound may suggest that an isosymmetric structural phase transition happens around 900 K \cite{Gibbs2011, Christy1995}.

\subsection{E. Bond lengths of Cr-O and Y-O}

To analyze detailed local crystalline environments of Cr and Y ions in the YCrO$_3$ compound, we extracted and plotted the lengths of Cr-O21, Cr-O22, and Cr-O1 [Fig.~\ref{Figure8}] and Y-O21, Y-O22, Y-O11, and Y-O12 [Fig.~\ref{Figure9}] bonds. The averaged Cr-O bond length, i.e., $<$Cr-O$>$, was calculated and plotted in addition [Fig.~\ref{Figure8}]. We tentatively refined the bond lengths with a piecewise linear function in the entire temperature regime with two ranges of 321--900 K and 900--1200 K. As shown in Figs.~\ref{Figure8} and \ref{Figure9}, the fit results (shown as solid lines) clearly display an anomaly around 900 K, consistent with our observation in the lattice configurations of \emph{a}, \emph{b}, \emph{c}, and \emph{V} [Figs.~\ref{Figure6} and \ref{Figure7}]. From the temperature range of 321--900 K to 900--1200 K, the slope of the bond length versus $T$ curve obviously  increases for the Cr-O21, Cr-O1, Y-O21, Y-O22, and Y-O12 bonds, e.g., for the Y-O22 bond, it increases from $4.42 \times 10^{-5}$ to $7.23 \times 10^{-5}$ \AA/K, by $\sim$ 63.57\%, whereas the slope sharply decreases for the Cr-O22 and Y-O11 bonds, e.g., for the Cr-O22 bond, it decreases by $\sim$ 62.59\%.

By comparing all Cr-O and Y-O bonds, we find that only the length of Cr-O21 bond increases in an inconspicuous way from 321 to 900 K, implying a small contribution to the thermal expansion of the lattice configuration. From 900 to 1200 K, the upturn of the lattice configuration (\emph{a}, \emph{b}, \emph{c}, and \emph{V}) was attributed to an increase of the lengths of Cr-O21, Cr-O1, Y-O21, Y-O22, and Y-O12 bonds. It is reasonable to deduce that the anisotropic thermal expansion in lattice constants \emph{a}, \emph{b}, and \emph{c} is from different increases in lengths along the different directions of Cr-O and Y-O bonds.

\begin{figure} [!t]
\centering
\includegraphics[width = 0.48\textwidth] {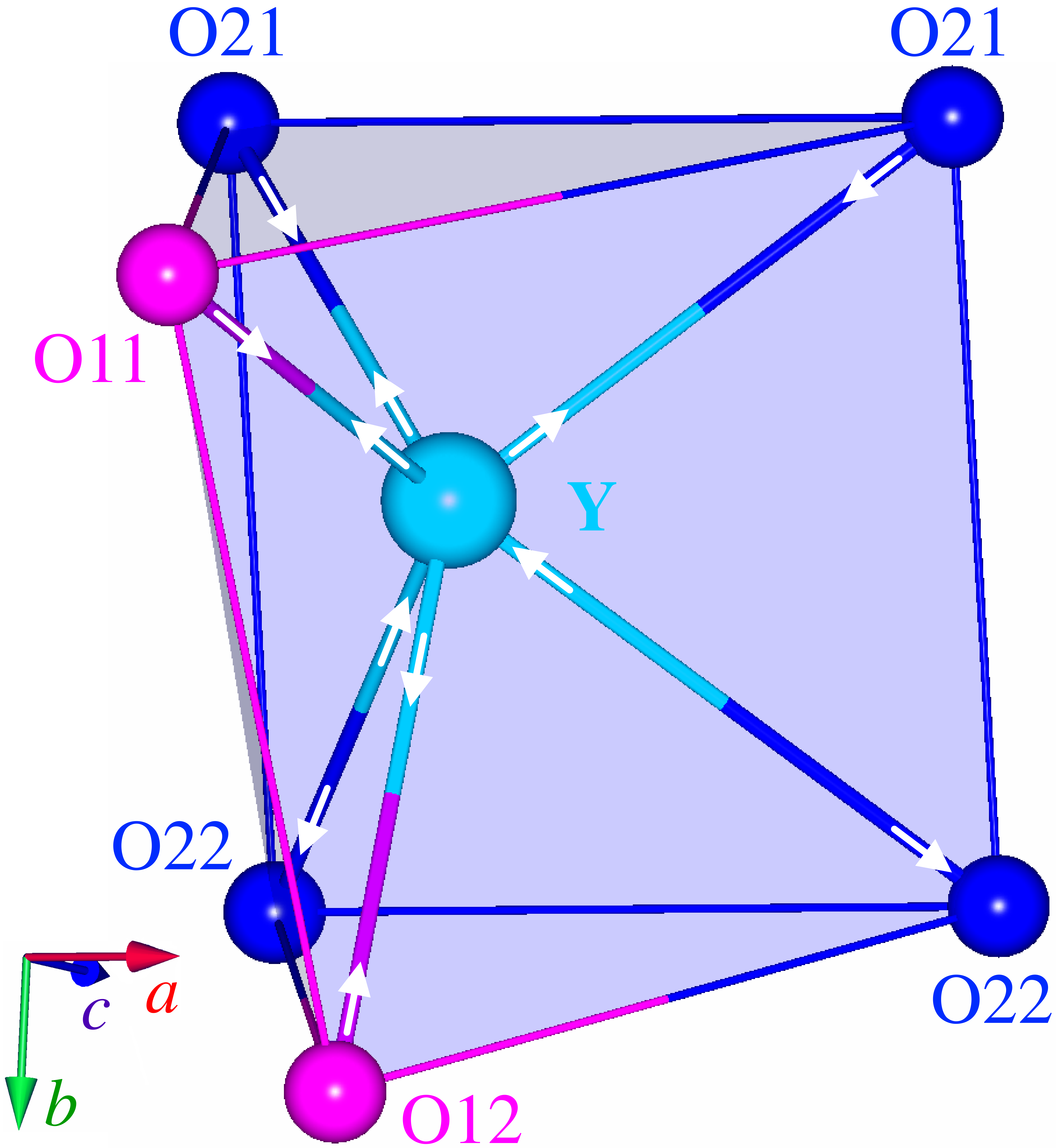}
\caption{Local pentahedron environment of Y ions in the single-crystal YCrO$_3$ compound, which was extracted based on our FULLPROF refinements \cite{Carvajal1993}. The Y, O11, O12, O21, and O22 ions are labeled as displayed. Detailed bond lengths of Y-O11, Y-O12, Y-O21 ($\times 2$), and Y-O22 ($\times 2$) are listed in Table~\ref{Table2}. The arrows sitting on the Y-O bonds schematically show the deduced pentahedron distortion configuration.}
\label{Figure11}
\end{figure}

\begin{figure*} [!t]
\centering
\includegraphics[width = 0.88\textwidth] {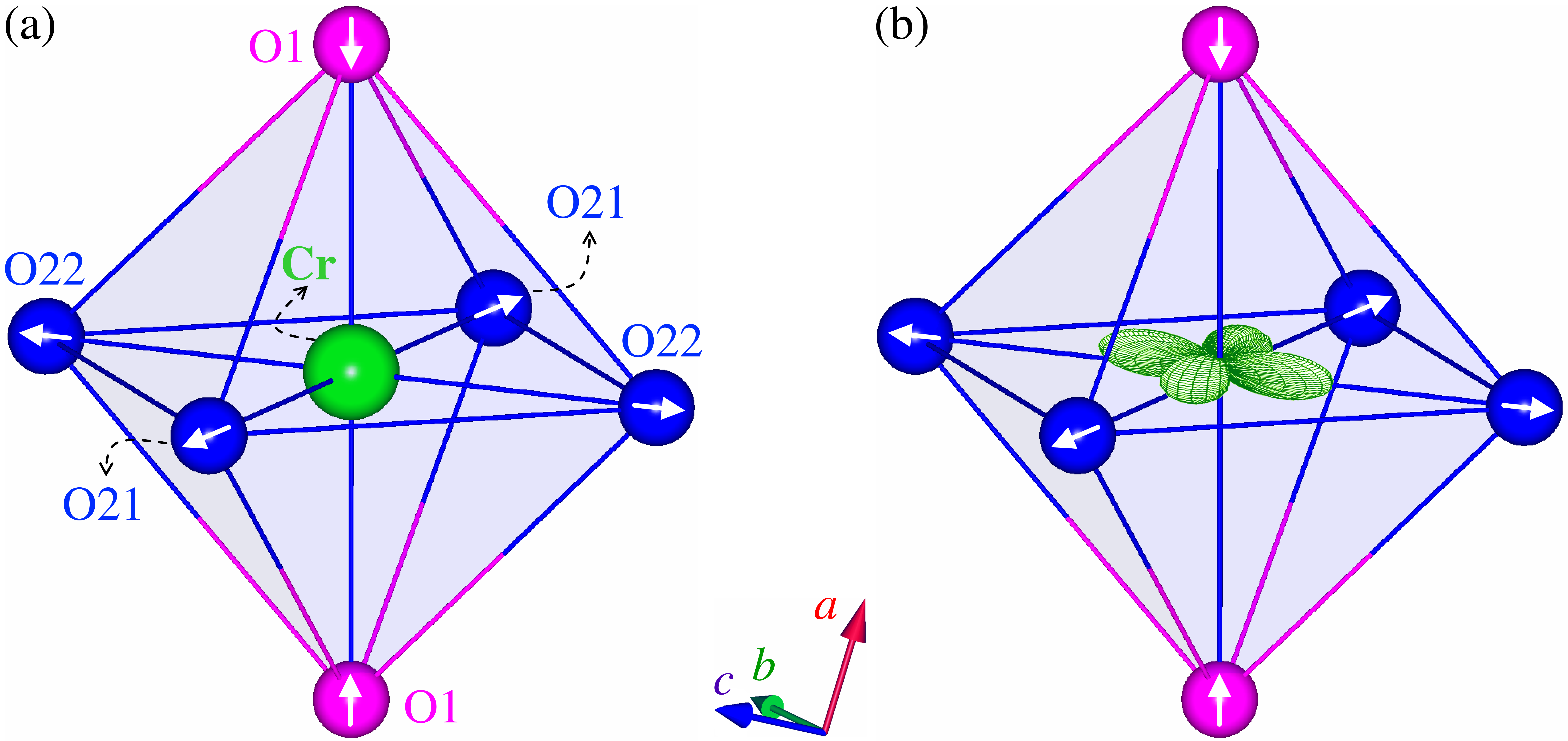}
\caption{(a) Local octahedral environment of Cr ion in the single-crystal YCrO$_3$ compound, which was extracted based on our FULLPROF refinements \cite{Carvajal1993}. The arrows drawn through the oxygen ions ($2\times$O1, $2\times$O21, and $2\times$O22) schematically show the deduced octahedral distortion mode. Representative refined bond lengths of Cr-O1, Cr-O21, and Cr-O22 at 321, 750, and 1200 K are listed in Table~\ref{Table2}. (b) In such octahedral geometry, we schematically drew the approximate 3$d_{yz}$ orbital shape in real space.}
\label{Figure12}
\end{figure*}

\begin{figure*} [!t]
\centering
\includegraphics[width = 0.88\textwidth] {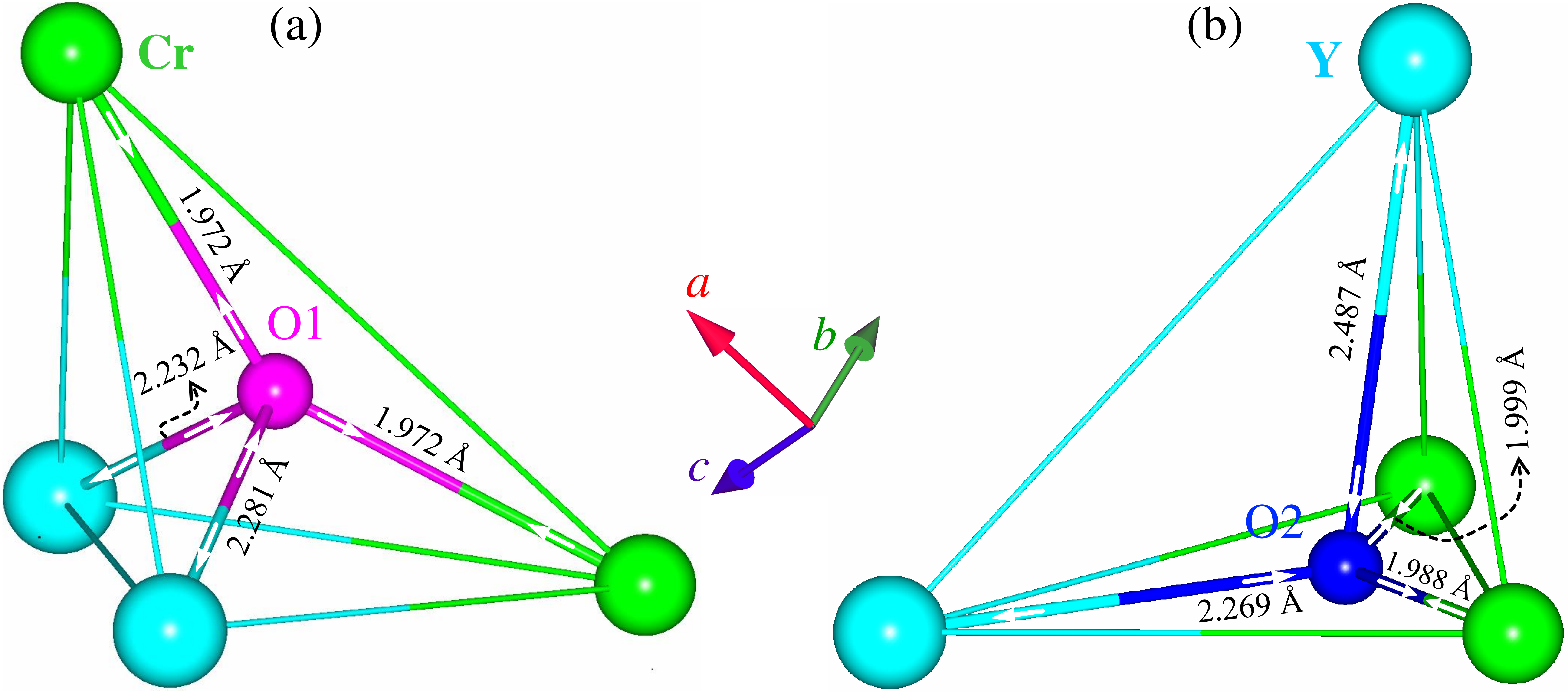}
\caption{Local tetrahedral environments of O1 (a) and O2 (b) ions in the single-crystal YCrO$_3$ compound, which was extracted based on our FULLPROF refinements \cite{Carvajal1993}. The Y, Cr, O1, and O2 ions are labeled as displayed. Detailed bond lengths of O1-Cr ($\times 2$), O1-Y, O2-Cr, and O2-Y at 321 K were marked. The arrows sitting on the O-Y and O-Cr bonds schematically show the deduced tetrahedral distortion modes.}
\label{Figure13}
\end{figure*}

\subsection{F. Local distortion modes of Y, Cr and O ions}

Identifying a detailed local crystalline environment is essential in determining electronic structure, spin configuration, orbital degeneracy and crystal field effect of 3\emph{d}-, 4\emph{d}-, 5\emph{d}-, or 4\emph{f}- compounds \cite{Li2008}. A quantitative measurement of the magnitude of a local crystalline environment can be evaluated by the local distortion parameter $\Delta$ defined as \cite{Li2008, Li2009, Li2007-1, Li2007-2}
\begin{eqnarray}
\Delta = \frac{1}{n}\sum_{i=1}^{n}\mathlarger{\mathlarger{\mathlarger{(}}}\frac{d_n - \langle d \rangle}{\langle d \rangle}\mathlarger{\mathlarger{\mathlarger{)}}}^2,
\label{distortion}
\end{eqnarray}
where \emph{n} is the coordination number, $d_n$ is the bond length along one of the \emph{n} coordination directions, and $\langle d \rangle$ is the averaged bond length. With Eq.~(\ref{distortion}), we calculated the local distortion parameter $\Delta$ as a function of temperature for the Y, Cr, O1, and O2 ions as shown in Fig.~\ref{Figure10}. For the 3\emph{d} Cr$^{3+}$ ions in single-crystal YCrO$_3$ compound, it is interesting that the local distortion parameter $\Delta$ keeps almost a constant in the whole temperature range, displaying no response to the previously reported dielectric anomaly \cite{Serrao2005}. The averaged $\Delta$ value of Cr$^{3+}$ ions from 321 to 1200 K is of $\sim$ $3.3 \times 10^{-5}$ that is approximately two orders of magnitude lower than that of 3\emph{d} Kramers Mn$^{3+}$ ions in the Jahn-Teller (JT) distorted regime of single-crystal La$_\frac{7}{8}$Sr$_\frac{1}{8}$MnO$_3$ compound \cite{Li2009}. This sharp comparison dramatically demonstrates the importance of local crystalline environment.

It is noted that the Y, O1, and O2 ions show huge $\Delta$ values, almost two orders of magnitude larger than that of Cr ions. At 321 K, $\Delta$(O2) $\approx$ 1.96$\Delta$(O1) $\approx$ 4.32$\Delta$(Y) $\approx$ 293.89$\Delta$(Cr), indicating a significant local crystalline environmental effect of O2, O1, and Y ions on the property of a YCrO$_3$ compound. This urges us to figure out the local distortion modes of Y, Cr, and O ions, as schematically drawn in Figs.~\ref{Figure11}--\ref{Figure13}, respectively.

Figure~\ref{Figure11} shows the local pentahedron environment of Y ions. It is worth mentioning that there are six Y-O bonds at the neighbor of the Y atom, i.e., with a coordination number 6, two Y-O1 and four Y-O2 bonds. Among them, 2$\times$Y-O22 bonds are stretched, and Y-O11, Y-O12, and 2$\times$Y-O21 bonds are shortened (Table~\ref{Table2}). This results in both the O11-O21-O22-O12 planes being bent outward, shifting the charge weight center of Y ions upward, whilst holding the 2$\times$O21 and the 2$\times$O22 ions within one O21-O22-O22-O21 plane. The extracted pentahedron distortion mode of Y ion was schematically displayed by the arrows sitting on the six Y-O bonds.

The Cr ion in YCrO$_3$ compound has three electrons in the unfilled 3\emph{d} shell, therefore, Cr$^{3+}$ is a non-Kramers ion, in principle, without JT effect. That is why Cr$^{3+}$ ions have a very small local distortion parameter $\Delta$ [Fig.~\ref{Figure10}]. From our refined Cr-O bond lengths as listed in Table~\ref{Table2}, we deduced the octahedral distortion mode of Cr$^{3+}$ ions (coordination number: 6) in YCrO$_3$ compound, as shown in Fig.~\ref{Figure12}(a). This distortion mode, as schematically displayed by the arrows sitting on oxygen ions, results from stretched Cr-O21 and Cr-O22 and shortened Cr-O1 bonds, behaving like a cooperative JT distortion. The local crystalline environment of Cr$^{3+}$ ions in YCrO$_3$ compound coincides with the $t_{\textrm{2g}}$ orbital shapes ($d_{xy}$, $d_{zx}$, $d_{yz}$). As shown in Fig.~\ref{Figure12}(b), the 3$d_{yz}$ orbital shape in real space was accommodated into the CrO$_6$ octahedron. Because the two Cr-O1/O21/O22 bonds locate along the plane/body diagonal direction and have the same length, the sum effect of electric-lattice interactions is canceled out. Therefore, within the present crystal symmetry, no displacement happens to the Cr ions.

Figure~\ref{Figure13} illustrates the local distortion environments of O1 [Fig.~\ref{Figure13}(a)] and O2 [Fig.~\ref{Figure13}(b)] ions in single-crystal YCrO$_3$ compound. Both O1 and O2 ions have a coordination number 4, forming a tetrahedron. By comparing the refined Cr-O and Y-O bond lengths (referring to Table~\ref{Table2}), we deduced two tetrahedral distortion modes for the O1 and O2 ions, respectively, as displayed by the arrows sitting on the Cr-O and Y-O bonds. For example, for the O1 ions [Fig.~\ref{Figure13}(a)], the 2$\times$Cr-O bonds are shortened, whereas the two Y-O bonds are stretched. This pushes the O1 ions towards to the Cr ions. So do the O2 ions [Fig.~\ref{Figure13}(b)]. Within the two oxygen tetrahedrons, the longest Y-O bond is Y-O2 = 2.487(4) {\AA} (at 321 K) as displayed in Fig.~\ref{Figure13}(b). This drives the O2 ions very close to the bottom Y-Cr-Cr plane, a large negative charge displacement. Compared to the O1 ions, the local crystalline environment of O2 ions is much more distorted in agreement with our calculations as shown in Fig.~\ref{Figure10} where the local distortion parameter $\Delta$ of O2 ions is approximately two times larger than that of O1 ions.

The distortion parameter $\Delta$, to some extent, is a criterion for what magnitude a certain atom displaces with its surrounding ligands \cite{Li2008}. For example, Kramers ions usually show a JT distortion whose magnitude can be expressed by the size of the distortion parameter $\Delta$. The JT effect occurring in 3\emph{d} transition metal oxides can lead to the degeneracy of \emph{d} orbitals accompanied by lowering the structural symmetry to release the electronic occupied energy. That can result in charge/orbital ordering and magnetic transition and may shed light on the colossal magnetoresistance effect \cite{Li2008, Millis1996}. For 3\emph{d}$^3$ Cr$^{3+}$ ions in YCrO$_3$ compounds, no JT effect is expected, and thus the distortion parameter $\Delta$ of Cr$^{3+}$ ions is quite small, which is hard to break the centrosymmetry of the center Cr$^{3+}$ ions, therefore, no ferroelectricity is expected from the Cr-sites. However, as shown in Figs.~\ref{Figure11} and \ref{Figure13}(b), the centrosymmetries of Y$^{3+}$ and O$_2^{2-}$ ions could be broken by their large local environmental distortions, which may induce the geometric ferroelectricity. Similar observation was also reported on the orthorhombic GdCrO$_3$ compound, in which the Gd$^{3+}$ and O$^{2+}$ ions move toward different directions and produce a huge charge density in the Gd-O bonds \cite{Mahana2018}.

\begin{figure} [!t]
\centering
\includegraphics[width = 0.48\textwidth] {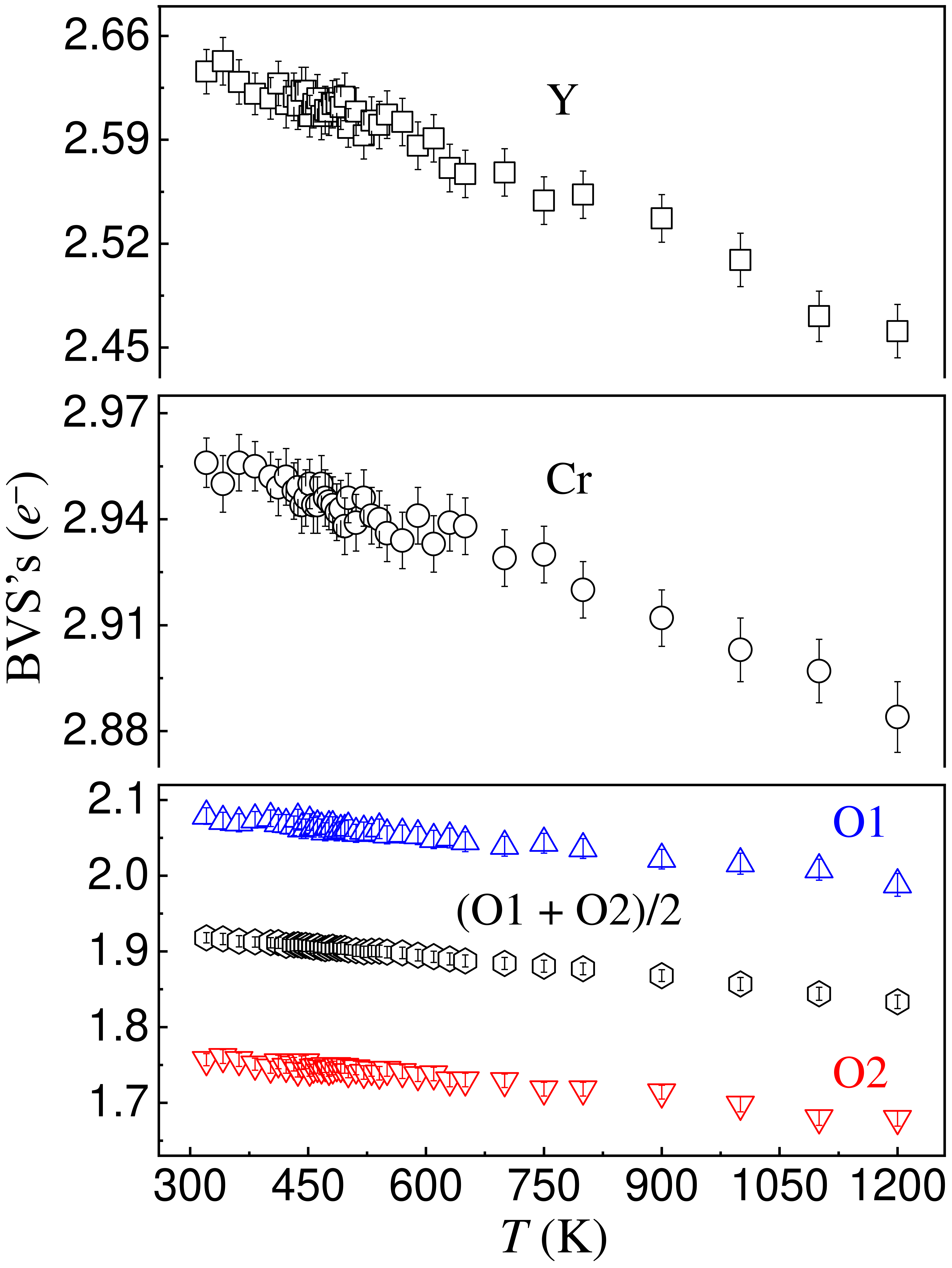}
\caption{Temperature variation of the bond valence states (BVSs) of Y, Cr, O1, and O2 ions in the single-crystal YCrO$_3$ compound, calculated from our refined structural parameters between 321 and 1200 K by the FULLPROF SUITE \cite{Carvajal1993}. For a clear comparsion, we also calculated the average BVSs of O1 and O2 ions, i.e., (O1 + O2)/2. Error bars are combined standard deviations.}
\label{Figure14}
\end{figure}

\subsection{G. Bond valence states of Y, Cr and O ions}

It is well known that bond valences are strongly correlated to bond distances, and some empirical relationships were previously proposed \cite{Carvajal1993, Brown1985, Brese1991, Mohri2000}. With our refinements, we extracted the bond valence states (BVSs) of Y, Cr, and O ions as shown in Fig.~\ref{Figure14}. As temperature increases, the BVSs of Y, Cr, and O ions almost decrease linearly. The calculated BVS values for the Cr and O1 ions are 2.956(7)+ and 2.079(11)-- at 321 K, close to the ideal 3+ and 2--, respectively. However, for the Y and O2 ions, the calculated BVS values, BVS(Y) = 2.636(15)+ and BVS(O2) = 1.757(8)--, differ largely from the respective perfect values of 3+ and 2-- in pure ionic model. Therefore, there exit positive and negative charge displacements from Y and O2 ions that coincide with their strongly distorted local crystalline environments. In the $Pmnb$ symmetry, there are two non-equivalent crystallographic sites for the O1 and O2 ions. As shown in Fig.~\ref{Figure14}, the charge difference between O1 and O2 ions is approximate 0.317(15)$e^{-}$ with $\sim$ 21 standard deviations, indicative of a big degree of the charge disproportion. It is of interest to notice that our extracted charge difference between O1 and O2 ions is even much bigger than the charge difference between Mn$^{3+}$ and Mn$^{4+}$ ions in the charge/orbital ordered states of manganites \cite{Li2009, Martin2004} where for the La$_\frac{7}{8}$Sr$_\frac{1}{8}$MnO$_3$ compound \cite{Li2009}, the charge difference is 0.11(5) $e^{-}$, and for the half-doped Nd$_\frac{1}{2}$Sr$_\frac{1}{2}$MnO$_3$ \cite{Martin2004} it is 0.16 $e^{-}$. It is thus reasonable to deduce that the Y and O2 ions play an important role in forming the dielectric anomaly of YCrO$_3$ compound.

\begin{figure} [!t]
\centering
\includegraphics[width = 0.48\textwidth] {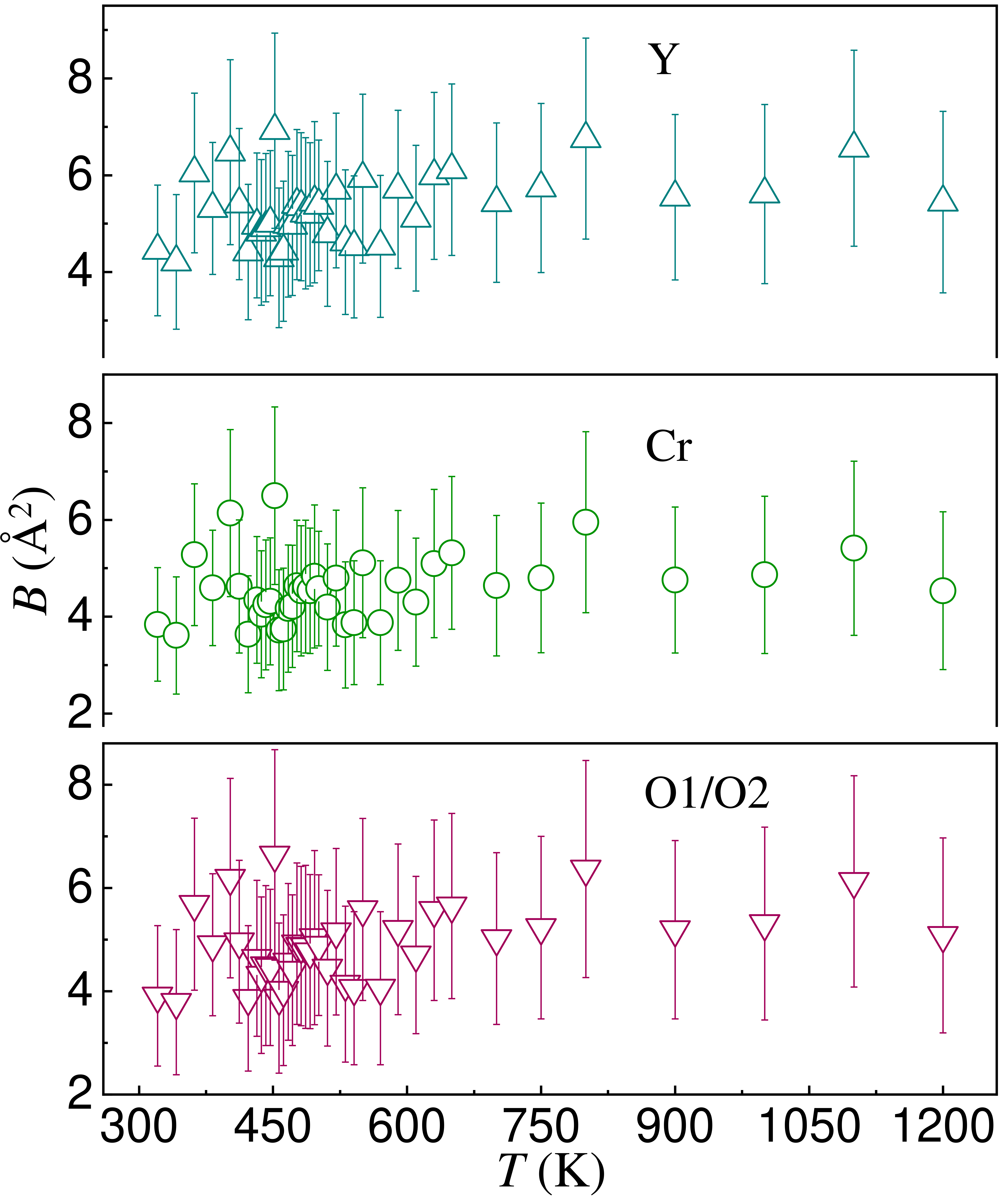}
\caption{Temperature variation of the isotropic thermal parameters, \emph{B}, of Y, Cr, and O1/O2 ions in the single-crystal YCrO$_3$ compound. During our FULLPROF \cite{Carvajal1993} refinements, we constrained the \emph{B} sizes of O1 and O2 ions to the same value. Error bars are standard deviations.}
\label{Figure15}
\end{figure}

Finally, the temperature-dependent isotropic thermal parameters $B$ of Y, Cr, and O1/O2 (constrained to be the same) ions of YCrO$_3$ compound were present in Fig.~\ref{Figure15} where almost no change exists from 321 to 1200 K within the experimental accuracy.

As is well known, the contribution of magnetism in YCrO$_3$ compound comes from the Cr$^{3+}$ (3\emph{d}$^3$) ions supposing that the YCrO$_3$ compound forms pure ionic bonds and thus there is no magnetic contribution from oxygen sites \cite{Li2008, Li2009}. As in the foregoing discussion, the dielectric anomaly \cite{Serrao2005} of YCrO$_3$ compound may be ascribed to the obvious atomic displacement and charge subduction of Y and O2 ions. The Y$^{3+}$ ion is nonmagnetic because the ground-state electronic configuration of neutral Y is [Kr]4\emph{d}$^1$5\emph{s}$^2$. Therefore, the structural parameters extracted from our time-of-flight neutron powder-diffraction study don't display a response to the previously observed dielectric anomaly. On the other hand, that doesn't mean that there is no magnetoelectric coupling existing in YCrO$_3$ compound. To unravel this coupling necessitates measurements with extremely high applied magnetic fields. By comparison, the YMnO$_3$ compound belongs to a hexagonal symmetry with the structural building block of MnO$_5$, and its ferroelectric polarization is due to opposing unequal dipoles of the two Y sites as well as the tilting and distortion of the MnO$_5$ blocks \cite{Aken2004, Fennie2005, Gibbs2011}. The Gd$^{3+}$ (4$f^7$) ions in GdCrO$_3$ compounds has a very strong magnetic contribution. It is expected that the Gd-Cr coupling in GdCrO$_3$ compounds is much stronger than the Y-Cr coupling in YCrO$_3$ compound. As foregoing remarks, the previously reported dielectric anomaly of YCrO$_3$ compound may be due to large atomic displacement and charge subduction of Y and O2 ions, whereas, only Cr ions contribute to its magnetism. That may be the reason why the structural parameters don't show any anomaly around 473 K.

It is stressed that the space group $Pmnb$ belongs to a centrosymmetric structure in which the sum of negative and positive charge shifts has to be zero. Our interesting observations, e.g., the extremely large local distortion parameter $\Delta$ of Y, O1, and O2 ions, the obvious displacement of Y and O2 ions and the charge subduction of Y and O2 ions, indicate that the actual structural symmetry of YCrO$_3$ compounds may be much lower than $Pmnb$. As a foregoing discussion, even though Kramers ions like Mn$^{3+}$ theoretically and experimentally exhibit a larger local distortion, e.g., a JT/cooperative JT effect, than non-Kramers ions like Cr$^{3+}$, their detailed distortion modes can be the same, as demonstrated in this study. Additionally this distortion mode can even be applicable to the 4$f$ ions \cite{HFLi2014}.

\section{IV. Conclusions}

To summarize, we have quantitatively investigated the high-temperature (300--980 K) magnetism and structural information (321--1200 K) of a single-crystalline YCrO$_3$ compound. The high-temperature magnetization can only be fit by the CW law with multiple effective PM moments and multiple PM CW temperatures with strong magnetic frustration, implying a complicated low-temperature magnetic structure. The magnetization versus applied-magnetic-field curve obeys well our modified Brillouin function with an inclusion of a factor $\eta$ denoting the strength of magnetic interactions. We refined the crystal structure with $Pmnb$ symmetry in the entire studied temperature range within the present experimental accuracy. Detailed structural information including lattice constants, unit-cell volume, atomic positions, thermal parameters, bond lengths, local distortion parameter, bond angles, local distortion modes, and BVSs, were extracted. The thermal expansions along the \emph{a}, \emph{b}, and \emph{c} axes are anisotropic with an anomaly appearing around 900 K. We attribute this anomaly to an isosymmetric structural phase transition. The local distortion parameter $\Delta$ of Cr ions is about two orders of magnitude lower than that of Y and O ions. We find that both Y and O2 ions produce a clear atomic displacement and a large charge deviation from theoretical ones. We thus suggest that the Y and O2 ions may play an important role in forming the previously reported dielectric anomaly of the YCrO$_3$ compound. The present results make YCrO$_3$ a particularly significant compound for theoretical and further experimental studies on $t_{\textrm{2g}}$ physics. The magnetic structure and dielectric property would be further explored in combination with theoretical calculations.

\section{Acknowledgements}

S.J. and D.-X.Y. acknowledge support from NKRDPC-2018YFA0306001, NKRDPC-2017YFA0206203, NSFC-11974432, NSFG-2019A1515011337, and Leading Talent Program of Guangdong Special Projects.
Z.T. acknowledges the start-up research grant (Grant No. SRG2016-00002-FST) at the University of Macau and the financial support from the Science and Technology Development Fund, Macau SAR (File No. 063/2016/A2).
D.O acknowledges financial support from the Science and Technology Development Fund, Macau SAR (File No. 0029/2018/A1).
H.{-}F.L acknowledges the start{-}up research grant (Grant No. SRG2016{-}00091{-}FST) at the University of Macau and financial support from the Science and Technology Development Fund, Macau SAR (File No. 064/2016/A2, File No. 028/2017/A1, and File No. 0051/2019/AFJ). This research used resources at the Spallation Neutron Source, a Department of Energy Office of Science User Facility operated by Oak Ridge National Laboratory.


\end{document}